\documentclass{aa}

\usepackage{graphicx}
\usepackage{txfonts}
\usepackage[]{hyperref}
\newcommand{\msun}{M_{\odot}}

\def\gapprox{\;\rlap{\lower 3.0pt                       
             \hbox{$\sim$}}\raise 2.5pt\hbox{$>$}\;}
\def\lapprox{\;\rlap{\lower 3.1pt                       
             \hbox{$\sim$}}\raise 2.7pt\hbox{$<$}\;}

\newcommand{\be}{ \begin{equation} }
\newcommand{\ee}{\end{equation}}

\newcommand{\ben}{\begin{enumerate}}
\newcommand{\een}{\end{enumerate}}

\usepackage{amstext}

\begin{document}

   \title{Impact of a star formation efficiency profile on the evolution of open clusters}
   \author{B.~Shukirgaliyev
          \inst{1,2}\fnmsep\thanks{Fellow of the International Max Planck Research School for Astronomy,
Cosmic Physics at the University of Heidelberg (IMPRS-HD)}
          \and
          G. Parmentier\inst{1}
          \and
          P. Berczik\inst{3,4,5}
          \and
          A. Just\inst{1}
          }

   \institute{Astronomisches Rechen-Institut, Zentrum für Astronomie der Universit\"at Heidelberg, M\"onchhofstr. 12-14, 69120 Heidelberg, Germany \\
              \email{beks@ari.uni-heidelberg.de}
         \and
             Fesenkov Astrophysical Institute, Observatory 23, 050020 Almaty, Kazakhstan\\
         \and
                         The International Center of Future Science of the Jilin University,
                         2699 Qianjin St., 130012, Changchun City, China\\
                 \and
             National Astronomical Observatories of China and Key Laboratory for
Computational Astrophysics, Chinese Academy of Sciences, 20A Datun Rd,
Chaoyang District, 100012 Beijing, China\\
         \and
             Main Astronomical Observatory, National Academy of Sciences of Ukraine,
27 Akademika Zabolotnoho St, 03680 Kyiv, Ukraine\\
             }

   \date{Received February 13, 2017 / Accepted June 05, 2017}


  \abstract
   {}
   {We study the effect of the instantaneous expulsion of residual star-forming gas on star clusters in which the residual gas has a density profile that is shallower than that of the embedded cluster.
   This configuration is expected if star formation proceeds with a given star-formation efficiency per free-fall time in a centrally concentrated molecular gas clump.}
   {
   We performed direct N-body simulations whose initial conditions were generated by the program "\textsc{mkhalo}" from the package "\textsf{falcON}",  adapted for our models.
   Our model clusters initially had a Plummer profile and are in virial equilibrium with the gravitational potential of the cluster-forming clump.
   The residual gas contribution was computed based on {a local-density driven clustered star formation model.}
   Our simulations included mass loss by stellar evolution and the tidal field of a host galaxy.}
   {We find that a star cluster with a minimum global star formation efficiency
(SFE) of 15 percent is able to survive instantaneous gas expulsion and to produce a bound cluster.
   Its violent relaxation lasts no longer than 20 Myr, independently of its global SFE and initial stellar mass.
   At the end of violent relaxation, the bound fractions of the
surviving clusters with the same global SFEs are similar, regardless of their initial stellar mass.
   Their subsequent lifetime in the gravitational field of the Galaxy depends on their bound stellar masses. }
   {We therefore conclude that the critical SFE needed to produce a bound cluster is 15 percent, which is roughly half the earlier estimates of 33 percent. 
   Thus we have improved the survival likelihood of young clusters after instantaneous gas expulsion.
   Young clusters can now survive instantaneous gas expulsion with a global SFEs as low as the SFEs observed for embedded clusters in the solar neighborhood (15-30 percent).
   The reason is that the star cluster density profile is steeper than that of the residual gas. 
   However, in terms of the effective SFE, measured by the virial ratio of the cluster at gas expulsion, our results are in agreement with previous studies. 
        }


   \maketitle
%
\section{Introduction} \label{sec:Intro}
The formation of bound star clusters can be divided into three phases: 1) star formation (SF), 2) expulsion of the residual star-forming gas, and 3) violent relaxation, that is, the cluster dynamical response to gas expulsion.
The dynamics of stars in young clusters during their formation and after gas expulsion is not fully understood yet.
It has been the object of intense scrutiny over the past years, with the applied methods ranging from pure N-body simulations \citep{Tutukov1978,Lada1984,Boily2003b, Baumgardt2007,Goodwin2009,Proszkow2009,Smith2011,Lee2016}
and combined hydrodynamical and N-body simulations \citep{Bonnell2011, Girichidis2012, Moeckel2012, Fujii2016} to analytical and semi-analytical models \citep{Hills1980, Adams2000, Boily2003a, PP2013}.

Star clusters form in dense (> 5\nobreakdash-10 $\times 10^3 \mathrm{cm}^{-3}$) clumps of gas inside giant molecular clouds (GMC) \citep{Lada2003, Lada2010, Kainulainen2014}.
The global star formation efficiency ($\mathrm{SFE_{gl}}$), the mass fraction of a star-forming region converted into stars, is defined as
\begin{equation}
\mathrm{SFE_{gl}} = \frac{M_\star}{M_\mathrm{gas}+M_\star},
\label{eq:SFE}
\end{equation}
where $M_\star$ is the total stellar mass and $M_\mathrm{gas}$ is the mass of unprocessed gas.

The star formation efficiencies (SFEs) measured from observations vary from a few to 30 percent for the dense clumps of molecular clouds \citep{Lada2003, Higuchi2009}, and from 0.1 percent to a few percent for their host GMCs \citep{Evans2009, Murray2011}.

Several mechanisms (stellar winds, ionizing radiation, radiation pressure, Type II supernova explosions) interrupt the SF process and blow up the unprocessed gas out of the cluster \citep{Krum2009,Murray2010,Dib2013,Hopkins2013}.
The observed open clusters older than 10 Myr are already gas
free \citep{Leisawitz1989,Lada2003}.
The duration of SF is on order of 1 Myr, with observations of young star clusters revealing stellar age spreads ranging approximately
between 0.3 Myr and 5.0 Myr \citep{Reggiani2011, Kudryavtseva2012}.

The combination of the SF duration with the SFE per free-fall time determines the global SFE achieved by a cluster-forming clump at the time of gas expulsion.
The SFE per free-fall time, the fraction of gas mass converted into stars over one free-fall time,  was estimated to be 0.01 by \citet{KrumholzTan2007}, while \citet{Murray2011} suggested
that it varies between 0.01 and 0.50 depending on the GMC mass.

\citet{Baumgardt2007} performed a grid of simulations assuming an embedded cluster in virial equilibrium with the residual gas,
{where both mass density profiles have identical shape.}
From these N-body simulations (see also Fig.~1 in \citet{Parmentier2007} for an overview of earlier works) it was concluded that a global SFE of at least 33 percent is needed to form a bound cluster after instantaneous gas expulsion.
This minimum global SFE is slightly higher than the SFEs observed for molecular clumps, which vary by up to 30 percent and are frequently estimated to be around 10 percent \citep{Higuchi2009,Murray2011,Kainulainen2014}.
To address this discrepancy between theoretical works and observations, different solutions exist: adiabatic gas expulsion { \citep{Lada1984,Geyer2001,Baumgardt2007,Brinkmann2017}, a subvirial cluster at the time of gas expulsion \citep{Verschueren1989, Goodwin2009, Farias2015}, or hierarchically formed clusters \citep{Smith2011,Lee2016}.}

\citet{Goodwin2009} stressed that the critical factor for a cluster to survive gas expulsion is its dynamical state (as measured by its virial ratio) at the onset of gas expulsion, and not its global SFE.
Because star clusters are not necessarily in equilibrium, he introduced an alternative SFE derived from the virial ratio of a star cluster measured immediately after instantaneous gas expulsion and called `effective SFE' (eSFE) (see also  \citet{Goodwin2006}):
\begin{equation}
\label{eq:eSFE}
\mathrm{eSFE} = \frac{1}{2 Q_\star},
\end{equation}
with the virial ratio defined by
\begin{equation}
\label{eq:Q_star}
Q_\star= \frac{T_\star}{|\Omega_\star|}.
\end{equation}
Here $T_\star$ and $\Omega_\star$ are the total kinetic and potential energies of a star cluster immediately after gas expulsion, and $Q_\star = 0.5$ corresponds to virial equilibrium.
The eSFE is equivalent to the global SFE for these models, where the SFE is constant with the distance to the center of the star-forming clump {(i.e., stars and gas follow the same density profiles shape)}, and stars are in virial equilibrium with the gravitational potential of residual gas.
According to \citet{Goodwin2009}, clusters whose virial ratio is $Q_\star < 1.5$ (equivalent to eSFE > 0.33) can survive the instantaneous gas expulsion.

\citet{Smith2011, Smith2013} and \citet{ Farias2015} studied this problem by proposing a hierarchical merging scenario of substructured embedded clusters.
They also concluded that the dynamical state of a cluster at the time of gas expulsion is important to cluster survival, while the global SFE is not.
However, their distributions of stars and star-forming gas are different, while not depending on each other.

The more recent work by \citet{Lee2016} proposed different dynamical states for the subclusters, which are in virial equilibrium with each other within the cluster-forming region.
The authors also concluded that the total dynamical state of the whole cluster at gas expulsion onset is the most important factor in predicting whether the cluster survives gas expulsion.
They did not link the formation of a bound cluster to its global SFE, however, and, considered only eSFEs.
One could nevertheless map their virial ratio to a subcluster mean SFE, assuming that in each subcluster, stars and gas present the same density profile shape. 

\citet{PP2013}, however, proposed a semi-analytical model of cluster formation in which the density profile of the embedded cluster is steeper than that of the cluster-forming gas.
That is, the SFE varies locally, as it steadily increases from the clump outskirts to the clump inner regions (see Fig. 10 in \citet{PP2013}).
The reason is that the cluster-forming clump is denser and therefore experiences faster SF in its central regions than in its outskirts.
The response to the gas expulsion of a cluster like this differs from the most often investigated case, that is, the case where the gas and stars density profiles have identical shapes. This was first investigated by \citet{Adams2000} with a semi-analytical method.
His choice for different gas and star density profiles was not physically motivated, however. \citet{Adams2000} showed that if the stellar mass is more concentrated in the cluster center than the gas mass, the cluster survival probability is significantly increased (see his Fig. 3).
The reason is a gas-poor region in the cluster central regions, which promotes the formation of a bound cluster, even when the global SFE is low. 
In addition, \citet{PP2014} performed N-body simulations whose initial conditions build on the model of \citet{PP2013} for a global SFE of about 18 percent.

In this contribution, we revisit this problem by means of direct N-body simulations.
Specifically, we build on the model of \citet{PP2013} to generate the initial conditions of our N-body simulations, thereby physically motivating our choice for the stars and gas density profiles of the embedded cluster at the time of the gas expulsion.
Our simulations also take into account stellar evolution and the tidal field of a Milky Way-like galaxy.

The outline of the paper is as follows.
In Sect. \ref{sec:SFE} we build on the model of \citet{PP2013} to relate the density profiles of the embedding gas and of the cluster at the onset of gas expulsion.
Section \ref{sec:tech} presents the technical setup and the initial conditions of our N-body simulations.

In Sect. \ref{sec:results} we discuss the results of our simulations
and compare them to earlier works.
Finally, we present our conclusions in Sect. \ref{sec:conc}.

\section{Star-formation efficiency and density profiles}\label{sec:SFE}
A semi-analytical model of star cluster formation from centrally
concentrated spherically symmetric gas clumps with a constant SFE per free-fall time, $\epsilon_\mathrm{ff}$, was developed by \citet{PP2013}.
Because the free-fall time is shorter in the clump inner (denser) regions than in the clump outer (less dense) regions, the density profile of the formed star cluster is steeper than the density profiles of the initial and  residual gas.
The authors considered that the total density profile $\rho_0$ of the system remains constant.

The total density profile is the sum of the density profiles of the embedded cluster, $\rho_\star$, and the residual gas, $\rho_\mathrm{gas}$, at any time $t$ after SF onset:
\begin{equation}
\label{eq.rhotot}
\rho_0(r) = \rho_\mathrm{gas}(t,r) + \rho_\star(t,r);
\end{equation}
here $r$ is the distance to the clump center.
The density profile of the unprocessed gas at time $t$ is described with Eq. (19) from \citet{PP2013}, which we reproduce here for the sake of clarity:
\begin{equation}
\label{eq.rhogas}
\rho_\mathrm{gas}(t,r)=\left( \rho_0(r)^{-1/2}+\sqrt{\frac{8G}{3\pi}} \epsilon_\mathrm{ff} t\right)^{-2}.
\end{equation}
$G$ is the gravitational constant and $\epsilon_\mathrm{ff}$ is the SFE per free-fall time. The mass of the embedded cluster at time $t$ is distributed according to Eq. (20) in \citet{PP2013}:
\begin{equation}
\label{eq.rhostar}
\rho_\star(t,r)=\rho_0(r) - \left( \rho_0(r)^{-1/2}+\sqrt{\frac{8G}{3\pi}} \epsilon_\mathrm{ff} t\right)^{-2}.
\end{equation}

Our aim is to investigate the dynamical evolution of such star clusters after instantaneous gas expulsion and to estimate their survival likelihood.
The instantaneous gas expulsion corresponds to the case when the timescale of the gas expulsion is significantly shorter than the dynamical timescale of the system.
In terms of cluster survivability, it is the worst scenario we can envision.
If a star cluster survives instantaneous gas expulsion, it is also able to survive a longer gas expulsion timescale.

We can adopt two different approaches to study this problem.

A. Either the starting point is a molecular clump, with a given density profile $\rho_0(r)$, and we obtain the density profile of the star cluster that formed after some SF time span based on \citet{PP2013}.

B. Alternatively, the starting point is an embedded cluster with a well-known profile (e.g., Plummer or King), and assuming an SF time span, we recover the initial gas density profile of the molecular clump out of which the cluster has formed.

In case (A) we start with the initial spherically symmetric gas clump that has a certain density profile $\rho_{0}(r)$ at time $t = 0$.
Then we assume that a star cluster forms within a time interval $ t = t_\mathrm{SF}$ called SF duration with a constant SFE per free-fall time $\epsilon_\mathrm{ff} = \mathrm{const}$.
Its density profile is then given by Eq. \ref{eq.rhostar}.
Depending on how long the SF process lasts, star clusters with different $\mathrm{SFE_{gl}}$ are formed (see Fig 9. in \citet{PP2013}).
At time $ t = t_\mathrm{SF}$ (corresponding to a certain value of  $\mathrm{SFE_{gl}}$), we set the instantaneous gas expulsion, that is, we remove the unprocessed gas from the system.
A star cluster then becomes super-virial because it has lost part of the gravitational potential within which it was in virial equilibrium.
This would define the initial conditions of our direct N-body simulations to study the effect of instantaneous gas expulsion.

Because model (A) depends on the SF duration, it results in star clusters with different global SFEs, masses, and spatial and velocity distributions, which makes it difficult to compare the results of the N-body simulations to each other.
Additionally, generating the initial conditions of such models for N-body simulations is not trivial.

Our main point is to study the effect of instantaneous gas expulsion on a star cluster in dependence of its global SFE, therefore
we can simplify the problem and consider clusters with a fixed stellar mass and spatial distribution while varying the global SFE at gas expulsion.
This leads us to our case (B), on which we focus in this paper.

In case (B), we thus assume a fixed density profile and a fixed stellar mass for the embedded clusters at gas expulsion.
Then the initial clumps that formed such clusters have therefore different total masses and spatial distributions depending on the assumed global SFE. 

As we wish to study the response of a star cluster to gas expulsion as a function of the global SFE, we have to find the cluster velocity distribution for any given global SFE assuming it is in virial equilibrium with the gravitational potential of the residual gas.
To find this distribution, which depends on the SF duration, we need to solve the inverse problem to that presented in case (A).
That is, having the density distribution of a star cluster $\rho_\star(r)$, we determine the density profile of the residual gas at gas expulsion $\rho_\mathrm{gas}(r,t_\mathrm{SF})$, and following from this, the density profile of the whole cluster-forming clump, $\rho_0(r,t_\mathrm{SF})$. Then we modify the cluster velocity distribution function so as to account for its virial equilibrium with the gravitational potential of the residual gas.

The model developed by \citet{PP2013} can be applied to any clump density profile. So we can choose a well-known density profile $\rho_\star$ for the embedded cluster and define the density profile of the residual gas corresponding to an SF duration $t_\mathrm{SF}$ by modifying Eqs. (\ref{eq.rhotot}) and (\ref{eq.rhogas}) and setting $t=t_\mathrm{SF}$:
\begin{equation}
\rho_0(r,t_\mathrm{SF}) = \rho_\mathrm{gas}(r,t_\mathrm{SF}) + \rho_\star(r),
\label{eq.rhotot1}
\end{equation}
\begin{equation}
\label{eq.rhogas1}
\rho_\mathrm{gas}(r,t_\mathrm{SF})=\left( \left(\rho_\mathrm{gas}(r,t_\mathrm{SF}) + \rho_\star(r)\right)^{-1/2}+\sqrt{\frac{8G}{3\pi}} \epsilon_\mathrm{ff} t_\mathrm{SF}\right)^{-2}.
\end{equation}
{In this equation, we note that the SF duration $t_\mathrm{SF}$ is given in units of Myr if densities $\rho_{(\star,~\mathrm{gas})}$ are expressed in $\msun~\mathrm{pc}^{-3}$  and the gravitational constant $G=0.0045~\mathrm{pc}^3~\msun^{-1}~\mathrm{Myr}^{-2}$.}
Introducing the parameter
\begin{equation}
\label{ka}
k = \sqrt{ \frac{8G}{3\pi} } \epsilon_\mathrm{ff} t_\mathrm{SF}
,\end{equation}
which depends on the SFE per free-fall time  $\epsilon_\mathrm{ff}$ and SF duration $t_\mathrm{SF}$ , we can rewrite Eq. (\ref{eq.rhogas1}) as
\begin{equation}
\label{polynom}
 k^4 \rho_\mathrm{gas}^4 -(4 k^2 - 2 k^4 \rho_{\star})\rho_\mathrm{gas}^3 -(6 k^2 \rho_{\star} - k^4 \rho_{\star}^2)\rho_\mathrm{gas}^2- 2 k^2\rho_{\star}^2 \rho_\mathrm{gas}+ \rho_{\star}^2 = 0.
\end{equation}
Solving this equation provides us with the residual gas density profile $\rho_\mathrm{gas}$ as a function of the stellar density profile $\rho_\star$, SFE per free-fall time , $\epsilon_\mathrm{ff}$, and SF duration $t_\mathrm{SF}$.
The stellar density profile, $\rho_\star$, can be any centrally
concentrated spherically symmetric density profile.
We find that two of the four roots of Eq. \ref{polynom} are complex. The other two roots are real, one increasing and one decreasing function of stellar density.
Since we consider a centrally concentrated clump, where the stellar density decreases with increasing radius, the gas density should follow the stellar density and decrease as well.
Thus we choose the root that is an increasing function of stellar density.
The mathematical expressions and a short analysis of these roots are provided in Appendix \ref{appenA}.

\citet{PP2013} used a power-law density profile with a slope of $-2$ for their cluster-forming clumps.
This yields a power-law density profile with a slope of about $-3$ for a star cluster.
Such initial conditions are not ideal for N-body simulations because of their infinite stellar and gas masses.
We need either to truncate these power-law profiles, or choose steeper profiles with finite masses.
Thus we decide to use one of the well-known spatial density distribution functions for an embedded cluster, that is, the Plummer profile \citep{Plummer1911},
\begin{equation}
\rho_{\star}(r) = \frac{3M_{\star}}{4\pi a_{\star}^3}\left(1+\frac{r^2}{a_{\star}^2}\right)^{-5/2}
\label{plummer}
,\end{equation}
where $M_\star$ is the cluster total mass and $a_\star$ is the Plummer radius, which corresponds to the projected half-mass radius of a star cluster.

Choosing a Plummer profile has many advantages, for instance, a finite mass for both gas and stars, and an analytical expression. It is also supported by almost all N-body codes, which makes it  possible to compare the results of different works.

\section{Technical setup} \label{sec:tech}
\subsection{Initial conditions} \label{sec:ICs}
Since our aim is to perform N-body simulations, we adopted dimensionless N-body ([NB]) units. {They are associated with the star cluster parameters, not with the cluster-forming clump parameters:}
\begin{eqnarray}
\label{eq:NBU}
G' = 1.0,\quad \quad
r' = \frac{r}{a_\star},  \label{eq.NBlen}\quad \quad
m' = \frac{m}{M_\star},  \label{eq.NBmass}\nonumber \\ 
v' = v\sqrt{\frac{a_\star}{G M_\star} },\quad \quad
t' = t \sqrt{\frac{G M_\star}{a_\star^3}}.
\end{eqnarray}
{The N-body units} can be converted into physical units when $G,\ a_\star,\ M_\star$ are assigned numerical values in their respective units. 
{Here we note that our `N-body' time unit depends on the cluster stellar mass $M_\star$ and Plummer radius $a_\star$ at the time of the instantaneous gas expulsion.
It does not represent the dynamical timescale of the cluster (stars only) because the cluster is super-virial and its dynamics bears the imprint of the cluster-forming clump mass at the time of the gas expulsion.
Neither does it represent the dynamical timescale of the clump (stars+gas) since the total mass and half-mass radius of the cluster-forming clump differ from the stellar mass and stellar half-mass radius because their density profiles have different shapes.
We applied these N-body units for (i) to recover the density profile of the residual cluster-forming gas with Eq. \ref{polynom}, and (ii) to perform the subsequent N-body integration of the gas-free cluster after instantaneous gas expulsion.
}

To generate the initial conditions of our N-body simulations is not trivial because we need a star cluster in virial equilibrium with the gravitational potential of the residual gas, where the shapes of the gas and star distributions differ.
In that respect, our case differs from most previous N-body simulations, as radial variations of the SFE increase the degree of complexity of the problem.

We used the \textsf{falcON} program \textsc{mkhalo} by \citet{McMillan2007}, which produces a spherically symmetric star cluster in virial equilibrium with an external potential as the initial conditions of direct N-body simulations.
External potential means a gravitational potential produced by anything but the stars of the cluster. In this framework, the gravitational potential produced by the residual gas constitutes an external potential.

To use \textsc{mkhalo} for our purpose, we wrote an additional acceleration plugin, that is, an additional code, which takes into account the new external potential of our models.
In this plugin we calculated the gravitational potential and the forces produced by the residual gas knowing its density profile. For this, we used Eq. (3.15$'$) from \citet{Duboshin1968}:
\begin{equation}
\label{D68:Phi}
\Phi_\mathrm{gas}(r) = - \frac{4 \pi G}{r}\int\limits_0^r r^2 \rho_\mathrm{gas}(r)dr - 4 \pi G \int\limits_r^{R_\mathrm{gas}} r \rho_\mathrm{gas}(r)dr;
\end{equation}
\begin{equation}
\label{D68:force}
\frac{d\Phi_\mathrm{gas}(r)}{dr} = \frac{4 \pi G}{r^2}\int\limits_0^r r^2 \rho_\mathrm{gas}(r)dr;
\end{equation}
where $\rho_\mathrm{gas}$ was obtained by solving Eq. \ref{polynom} (see Sect. \ref{sec:SFE}), and $R_\mathrm{gas}$ is the adopted outer edge of the clump of residual gas. We used $R_\mathrm{gas} = 32 a_\star$, which is the smallest radius possible to use in \textsc{mkhalo}.
Because the gas density profile, $\rho_\mathrm{gas}$, is not a simple function of $r$, the distance to the clump center, Eqs. (\ref{D68:Phi}) and (\ref{D68:force}) were integrated numerically using the Simpson method.

In the framework of this study, we adopted a fixed SFE per free-fall time of $\epsilon_\mathrm{ff} = 0.05$.
Then the only parameter we varied in our acceleration plugin to produce the initial conditions is the SF duration $t_\mathrm{SF}$.
To allow a comparison of our results with other works, however, it is better to use as a main parameter the $\mathrm{SFE_{gl}}$  than $t_\mathrm{SF}$.
To develop a grid of models with a given $\mathrm{SFE_{gl}}$ (0.05, 0.10, ...), we still need to infer the corresponding $t_\mathrm{SF}$ and add them to the models.

We defined the global SFE as the ratio between the stellar and total (stellar + gas) masses residing inside a chosen outer limit, $R_{cl}$.
Because a Plummer model has no finite outer limit, we adopted $R_{cl} = 10 a_\star$ , which is the radius inside which about 98 per cent of the stellar mass resides.
Here we note that because of the slope difference between the density profiles of the embedded cluster and the residual (as well as total) gas, a larger outer limit would imply a lower global SFE, and vice versa.

\begin{figure}
\centering
\includegraphics[width=\hsize]{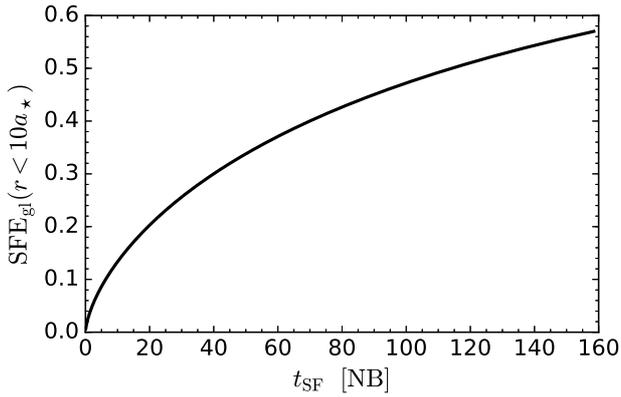}
\caption{Relation between the SF duration $(t_\mathrm{SF})$ {(given in N-body units according to Eq.\ref{eq:NBU})} and the {global SFE (}$\mathrm{SFE_{gl}}$), measured inside $R_{cl}=10a_\star$. The lowest value of the global SFE here in this plot is $\mathrm{SFE_{gl}}=0.007$, which corresponds to an SF duration $t_\mathrm{SF} = 0.01$~[NB].}
\label{fig3:SFE_gl}
\end{figure}

The relation between the $\mathrm{SFE_{gl}}$ and the corresponding SF duration $t_\mathrm{SF}$ calculated in N-body units is presented in Figure \ref{fig3:SFE_gl}. 
In our study we concentrated on models with an $\mathrm{SFE_{gl}} < 0.50$.
The corresponding values of $\mathrm{SFE_{gl}}$ and $t_\mathrm{SF}$ for different models are presented in Table \ref{tab:tsfsfe}.
Using these values, we produced the initial conditions of our simulations with \textsc{mkhalo}.
Then, using the generated positions and velocities of the stars, we calculated the initial potential ($\Omega_\star$) and kinetic ($T_\star$) energies of our model clusters at the moment of instantaneous gas expulsion, as well as their initial virial ratios and effective SFEs,  using Eqs. (\ref{eq:Q_star}) and (\ref{eq:eSFE}).
The corresponding values are also presented in Table \ref{tab:tsfsfe}.

\begin{table}
  \caption{ {Global SFEs} ($\mathrm{SFE_{gl}}$) and their corresponding SF durations ($t_\mathrm{SF}$) in {N-body units (see Eq. \ref{eq:NBU})}, the virial ratios $Q_\star$ of star clusters, and effective SFEs  (eSFE) immediately after gas expulsion.}
\label{tab:tsfsfe}
\centering
  \begin{tabular}{ l r r  r}
  \hline\hline
$\mathrm{SFE_{gl}}$ &  $t_\mathrm{SF}$ [NB]  & $Q_\star = T_\star/|\Omega_\star|$ & $\mathrm{eSFE}=1/2Q_\star$\\
  \hline
0.05 &  2.14   & 4.34 & 0.12\\
0.10 &  6.30   & 2.26 & 0.22\\
0.13 &  9.58   & 1.77 & 0.28\\
0.15 &  12.09  & 1.55 & 0.32\\
0.20 &  19.53  & 1.21 & 0.41\\
0.25 &  28.74  & 1.00 & 0.50\\
0.30 &  39.96  & 0.87 & 0.58\\
0.35 &  53.53  & 0.77 & 0.65\\
0.40 &  69.94  & 0.71 & 0.70\\
0.45 &  89.85  & 0.66 & 0.76\\
0.50 &  114.22 & 0.62 & 0.81\\
\hline
  \end{tabular}
\end{table}
We find that in our models the effective and global SFEs are different, unlike the models that used the same density profile for both the stars and gas (see \citet{Boily2003a}, \citet{Goodwin2006}, \citet{Goodwin2009}).
Varying the outer edge $R_{cl}$ of the cluster, we infer that these eSFEs are almost the same as the global SFEs measured inside $1.5a_\star$ , that is, similar within 2-3 percent to the global SFEs measured inside a cluster half-mass radius.
The latter also measures the local stellar fraction (LSF) as defined by \citet{Smith2011}.
\citet{Goodwin2009} noted that star clusters with an initial virial ratio lower than  1.5 are able to survive the instantaneous gas expulsion (in the case of Plummer profiles for both stars and gas), which corresponds to an effective SFE of 33 percent.
Taking this into account, we can expect the minimum SFE needed to form a bound cluster to be $\mathrm{SFE_{gl} = 0.15}$, as it corresponds to $Q_\star \approx 1.5$ for our models (see Table \ref{tab:tsfsfe}).
This is indeed what we show in Sect. \ref{sec:results}.

\begin{figure}
\centering
\includegraphics[width=\hsize]{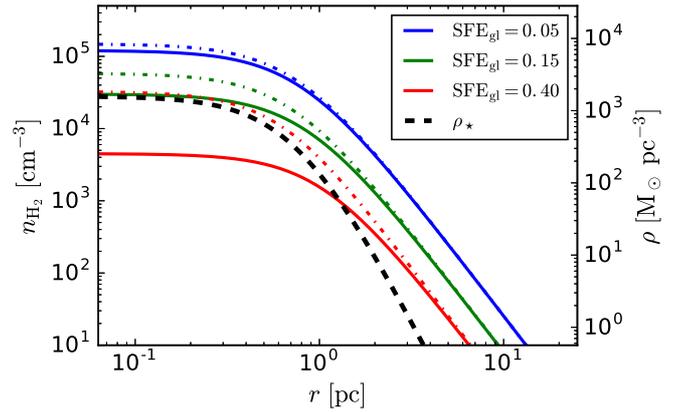}
\caption{Density profiles of the star cluster (black dashed line), of the residual (solid lines), and initial (dash-dotted lines) gas for different $\mathrm{SFE_{gl}}$ in scaled physical units. A total stellar mass $M_\star = 3000\ \msun$ and a 3D half-mass radius $r_h = 1 \mathrm{\ pc}$ are assumed. Note that the stellar density profile is a Plummer profile.}
\label{fig1:dens_prof}
\end{figure}

The density profiles of the embedded cluster, its residual gas before instantaneous gas expulsion, and the initial clump gas for different global SFEs are shown in Figure \ref{fig1:dens_prof}.
The models were scaled to physical units assuming a star cluster mass of $M_\star = 3000\ \msun$ and a 3D half-mass radius of $r_h=1 \mathrm{\ pc}$.
The density units are given in $\msun\/\mathrm{pc^{3}}$ (right y-axis) and molecules per cm$^{3}$ (left y-axis).
Gas densities on this scale vary within the observed range of dense clumps ($\lapprox 10^5 \mathrm{cm}^{-3}$).
The SF durations $t_\mathrm{SF}$ for these three models are 0.39 Myr, 2.21 Myr, and 12.77 Myr.
As we see, some of our models are inconsistent with observations in the sense that the inferred SF duration is longer than what is observed.

For our high-resolution direct N-body simulations we have chosen the \textsc{$\phi$grape-gpu} code developed by \citet{Berc2011}.
As a check of the stability of the initial conditions generated by the code \textsc{mkhalo}, we added our newly created external potential to the \textsc{$\phi$grape-gpu} code and tested the dynamics of the embedded cluster within the residual gas potential.
We ran a few simulations with external potentials corresponding to different $t_\mathrm{SF}$.
The test runs were performed for isolated clusters with $N=10$k single-mass particles over a time interval of up to 1000 N-BODY time units, which is slightly shorter than three relaxation times of the same cluster in virial equilibrium without external potential.
We checked the evolution of the Lagrangian radii and cumulative mass profiles for two models  with low and high values of $t_\mathrm{SF}$.
These exercises showed that our newly generated initial conditions are indeed in virial equilibrium with the external gas potential.

\subsection{Setting up the simulations}

We performed two types of N-body simulations. First, we simulated isolated single-mass clusters without stellar evolution to observe the pure dynamical effect of an instantaneous gas expulsion. 
We ran these simulations with $N=10^4$ particles and covered global SFEs ranging from 5 to 50 percent.

Second, we studied a more realistic scenario of violent relaxation by considering star clusters consisting of multi-mass stars that evolve in a Galactic tidal field.
We refer to these two sets of simulations as `isolated' and `non-isolated' models.

For the stellar initial mass function (IMF) of our non-isolated models we adopted the IMF of \citet{K2001} with the lower and upper mass limits of $M_\mathrm{low} = 0.08\ M_\odot$ and $M_\mathrm{up} = 100\ M_\odot$, respectively.

To account for the tidal field of the Galaxy, we used an axisymmetric three-component model that consists of a bulge, disk, and halo, as described by the Plummer-Kuzmin model \citep {Miyamoto1975}. This was added to the \textsc{$\phi$grape-gpu} code by \citet{Just2009}, and we used it keeping their parameters.

We considered that our clusters move on a circular orbit in the plane of the Galactic disk, at a distance of $R_0 = 8\ \mathrm{kpc}$ from the Galactic center.

We normalized our N-BODY units to the real physical units in order to assign the correct timescale to the stellar evolution routines (SSE; \citet{Hurley2000}) implemented in the \textsc{$\phi$grape-gpu} code. 
This means that we assigned certain values to the cluster mass ($M_\star$) and the initial Plummer radius ($a_\star$).

Our simulations encompass five different initial cluster stellar masses: $M_\star=3000$,\ 6000,\ 10000,\ 15000, and $30000\ M_\odot$.
Then knowing the distance of the cluster to the Galactic center, we can calculate the cluster tidal (Jacobi) radius $r_t$ for a given mass $M_\star$ using Eq (13) from \citet{Just2009}, which we reproduce here for the sake of clarity:
\begin{equation}
\label{rtid}
r_t = \left( \frac{G M_\star}{(4-\beta^2)\Omega^2} \right)^{1/3}.
\end{equation}
Here $\beta=1.37$ is the normalized epicyclic frequency and $\Omega = V_0/R_0$ is the angular speed of a star cluster on a circular orbit at a distance $R_0$ from the Galactic center.
For $R_0 = 8000 \mathrm{\ pc}$, the orbital speed recovered from the rotation curve of the Galaxy model provided in \citet{Just2009} is $V_0 = 234.24\mathrm{\ km\ s^{-1}}$.

To make the models comparable with each other, we fixed the half-mass radius to the tidal radius ratio.
We calculated it for a cluster mass $M_\star = 3000 \msun$ with a half-mass radius of $r_h = 1\ \mathrm{pc} $.
This means that we considered clusters with different stellar masses, but the same mean stellar volume densities.
The tidal radius of a cluster with $M_\star = 3000 \msun$ and $r_h = 1\ \mathrm{pc} $ is $r_t = 19.2$ pc, and therefore
\begin{equation}
\label{ratio}
\frac{r_h}{r_t} \approx 0.052.
\end{equation}

Thus we normalized the N-BODY length unit into physical units of pc knowing that in a Plummer model $ r_h \approx 1.3 a_\star $,
\begin{equation}
\label{rnorm}
r_\mathrm{norm} = a_\star \approx 0.77 r_h \approx 0.04 r_t
,\end{equation}
where the tidal radius is calculated using Eq. (\ref{rtid}).
For the models with an initial stellar mass $M_\star = 3000\ M_\odot$ , for instance, the normalized length is equal to
$
r_\mathrm{norm} =  0.77\ \mathrm{pc}.
$

With our definition of the outer limit $R_{cl}$ of our cluster-forming clumps (see Sect. \ref{sec:ICs}), star clusters initially fill their tidal radius up to 40 percent. This means that the total radius of a star cluster is initially smaller than its tidal radius: $R_\mathrm{cl} = 10a_\star = 0.4r_t$.  The properties of a star cluster as a function of its stellar mass are presented in Table \ref{tab:M}.
\begin{table}
\caption{Set of models corresponding to different initial stellar masses.}
\label{tab:M}
\centering
\begin{tabular}{ r r r r r }
\hline\hline
$M_\star$  & $N_\star$ & $r_t$ & $r_h$ & $R_\mathrm{cl} = 10a_\star$  \\
$[M_\odot]$ &          & [pc]             &  [pc]             & [pc] \\
\hline
 3000 &  5227 & 19.211 & 1.00 & 7.664\\
 6000 & 10455 & 24.204 & 1.26 & 9.656\\
10000 & 17425 & 28.697 & 1.49 & 11.449 \\
15000 & 26138 & 32.850 & 1.71 & 13.310\\
30000 & 52277 & 41.389 & 2.15 & 16.512 \\
\hline
\end{tabular}
\end{table}

The scale factor of time units,{ as shown in Eq. \ref{eq:NBU},} can be found as
\begin{equation}
\label{tnorm}
t_\mathrm{norm} = \sqrt{\frac{r_\mathrm{norm}^3}{G M_\star}}\approx0.18\ \mathrm{Myr}\end{equation}
when $r_h/r_t=0.052$. 
Given our assumption of a fixed ratio of the half-mass to tidal radius (see Eqs. \ref{ratio} and \ref{rnorm}), $r_\mathrm{norm} \propto r_t \propto (GM_\star)^{1/3}$, $t_\mathrm{norm}$ is the same for all models as well as the mean stellar volume densities.

The corresponding values of SF duration and total (stars + gas) volume densities averaged within the initial stellar 3D half-mass radius are provided in Table~\ref{tab:tsf}.
This table shows that the models are consistent with the observations in terms of SF duration and clump mean densities.
It is thought that star cluster formation takes between one half and roughly 5 Myr in the solar neighborhood.
We therefore consider the models in these limits to make our simulations as consistent with reality as possible.
This limits the SF duration of our models between 2.7 and 27 N-BODY time units, which corresponds to 0.5 and 5 Myr for our chosen scale factor of ${r_h}/{r_t}$.
Consequently, this also limits us in the range of achievable $\mathrm{SFE_{gl}}$.
According to these limits, we decided to calculate models with $\mathrm{SFE_{gl}}$ between 10 percent and 25 percent for $\epsilon_\mathrm{ff} = 0.05$, which corresponds to an SF duration $t_\mathrm{SF}$ between 1.15 and 5.25 Myr for all initial cluster masses $M_\star$.
To cover still higher SFEs, we ran two additional models with global SFEs of 30 and 35 percent for $M_\star = 6000 \msun$. 
To make these runs consistent with the observations in terms of SF duration, we built on the following feature of Eq. (\ref{polynom}):
because the parameter $k$ in Eq. (\ref{polynom}) is proportional to $\epsilon_\mathrm{ff} t_\mathrm{SF}$, our results stand for any model where the product $\epsilon_\mathrm{ff} t_\mathrm{SF}$ is conserved (e.g., a twice higher SFE per free-fall time with a twice shorter SF duration).
In our two additional runs the SF durations can therefore be considered as $t_\mathrm{SF} = 3.64 $ Myr and 4.88 Myr, respectively, if $\epsilon_\mathrm{ff} = 0.1$.
For comparison, additional models with the same spatial distribution of stars initially in virial equilibrium  within a Galactic tidal field, but without any residual star-forming gas, (i.e., equivalent to $\mathrm{SFE_{gl}}=1.0$), were run for $M_\star = $ 3000, 6000, and 
10000 $\msun$.

The mean (total and stellar) volume densities of models are also consistent with observations of star-forming molecular clumps, {where an SF density threshold has been suggested} ($\gapprox 10^4\ \mathrm{cm}^{-3}$ in \citet{Lada2010} and $>5\times10^3\ \mathrm{cm}^{-3}$ in \citet{Kainulainen2014}) and with stellar densities in embedded clusters, which vary from 100-200 to 1-2$\times 10^4$ stars~pc$^{-3}$ \citep{Lada1991,Hillenbrand2000}.
\begin{table}
  \caption{ {Global SFEs} ($\mathrm{SFE_{gl}}$) and corresponding values of the SF duration ($t_\mathrm{SF}$) in {N-body units (Eq. \ref{eq:NBU})} and in units of Myr for the adopted normalization parameters {(using Eq. \ref{tnorm})}. {The rows within the box highlight the non-isolated models selected for our simulations.} We also show here the mean volume density of the clump (star + gas) inside the half-mass radius. Because we kept the stellar masses constant while varying the global SFE, we varied the fraction of unprocessed gas. When the global SFE increases, the mass of the residual gas decreases (keeping stellar mass constant), and as a consequence, the total volume density decreases as well.}
\label{tab:tsf}
\centering
  \begin{tabular}{ r r r r r}
  \hline\hline
$\mathrm{SFE_{gl}}$ &  $t_\mathrm{SF}$ & $t_\mathrm{SF}$ & $\overline\rho_\mathrm{tot}(<r_h)$& $\overline n_\mathrm{H_2,tot}(<r_h)$\\
& [NB] & [Myr] & $[ M_\odot\ \mathrm{pc}^{-3}]$ & $[ \mathrm{cm}^{-3}]$\\
  \hline
0.05 &  2.14       &  0.39 &\\ \cline{1-5}
\multicolumn{1}{|r} {0.10} &  6.30       &  {1.15} & 1490.73& \multicolumn{1}{r|}{26405}\\
\multicolumn{1}{|r} {0.13} &  9.58       &  1.75 & 1177.02& \multicolumn{1}{r|}{20848}\\
\multicolumn{1}{|r} {0.15} &  12.09      &  2.21 & 1037.99& \multicolumn{1}{r|}{18386} \\
\multicolumn{1}{|r} {0.20} &  19.53      &  3.56 & 813.52& \multicolumn{1}{r|}{14410} \\
\multicolumn{1}{|r} {0.25} &  28.74      &  5.25 & 680.92& \multicolumn{1}{r|}{12061} \\ \cline{1-5}
0.30 &  39.96      &  7.29 & 594.24&  10526\\
0.35 &  53.53      &  9.77 & 533.91&  9457\\
0.40 &  69.94      & 12.77 &\\
0.45 &  89.85      & 16.40 &\\
0.50 &  114.22     & 20.85 &\\
\hline
  \end{tabular}
\end{table}

Based on these simulations, we now study the evolution of the bound fraction of star clusters.
We present the results of isolated models scaled to physical units such that $M_\star~=~6000~\msun$ and $r_h = 1.26$~pc (see Table \ref{tab:M}) {to compare them with $M_\star~=~6000\ \msun$ non-isolated models, which consist of 10455 stars. }

\section{Results and discussions}\label{sec:results}

\subsection{Cluster-bound fraction evolution}
{
In the classical way,} the bound fraction of isolated clusters in virial equilibrium is defined as the fraction of stars whose total energy is negative, that is, the potential energy dominates the kinetic energy.
Our model isolated clusters become supervirial and start to expand quickly after instantaneous gas expulsion, however.
Thus it is not trivial to distinguish between bound and unbound stars in such systems during their expansion.
This process can take quite a long time in isolated systems, especially if the final bound fraction is small. 
For instance, the cluster with $\mathrm{SFE_{gl}}=0.15$ needs 1 Gyr to approach the final bound fraction if this is defined in the ``classical'' way. 
We therefore developed the following technique to determine the final bound fraction.
We recalculated the total energies of stars, removing the unbound stars (i.e., stars whose total energy is positive) from the cluster even if they were located in the center of the cluster. We iterated until no unbound star remained in the cluster. 
Doing so, we excluded the contribution of unbound stars to the cluster potential.
The final bound fraction of our isolated models is defined as the final fraction of stars that remained bound to the cluster - if any - at the end of the iterations.
We emphasize that removing the unbound stars does not mean that we removed them from the simulations, but that we removed them from our selected sample at each snapshot for the purpose of our analysis only. 
For each snapshot in time, we always started the iterative process with the total number of stars.
More discussion about this technique and its relevance can be found in Appendix \ref{AppenB}. 

{For the non-isolated clusters, that is, those evolving within a Galactic tidal field, the bound fraction is defined as the stellar mass residing inside the instantaneous tidal radius normalized to the initial stellar mass,}
\begin{equation}
\label{eq:fbound}
F_{bound} = \frac{M_\star(r<r_t)}{M_\star}.
\end{equation}

\begin{figure}[!h]
\centering
\includegraphics[width=1.05\columnwidth]{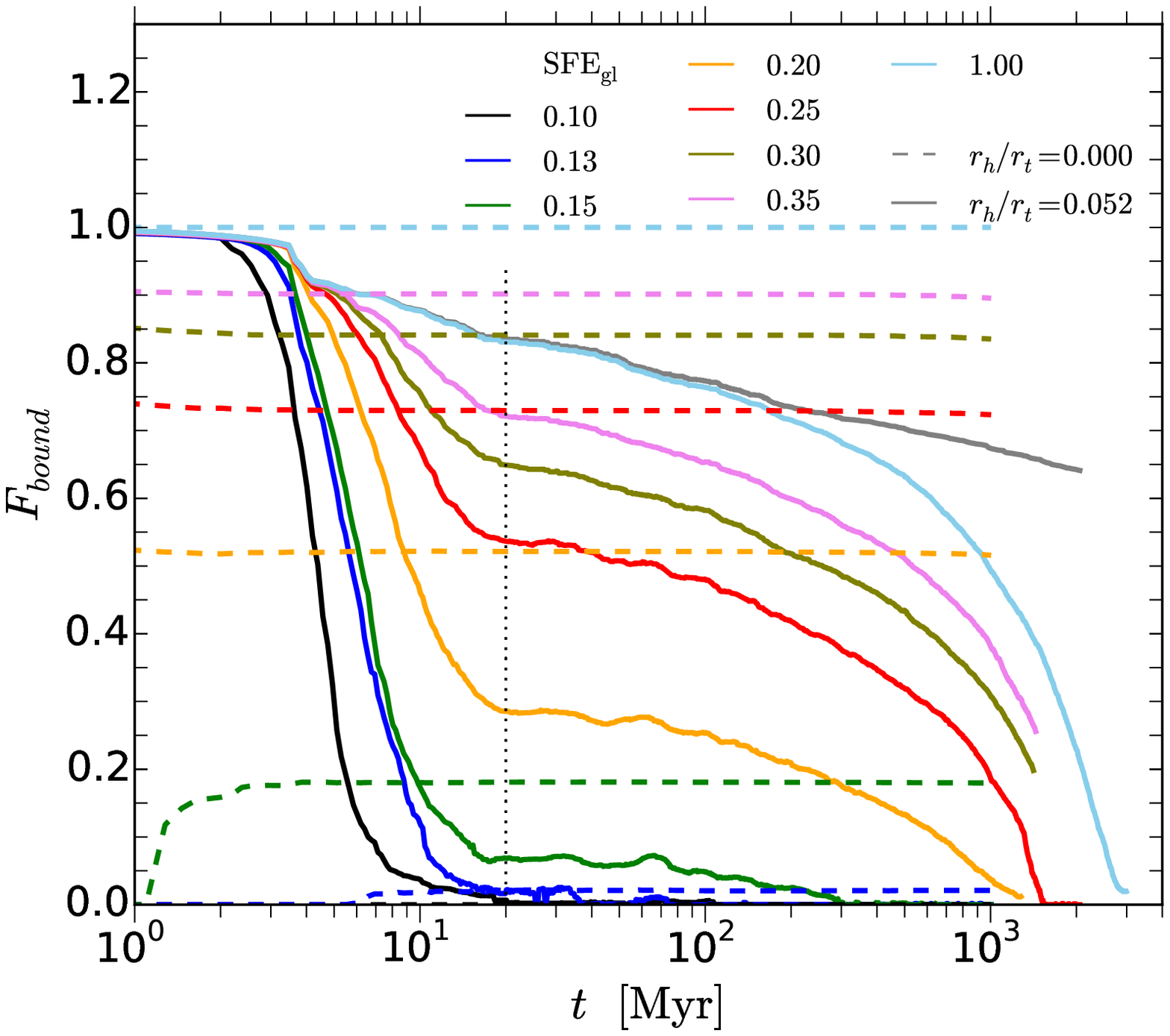}
\includegraphics[width=1.05\columnwidth]{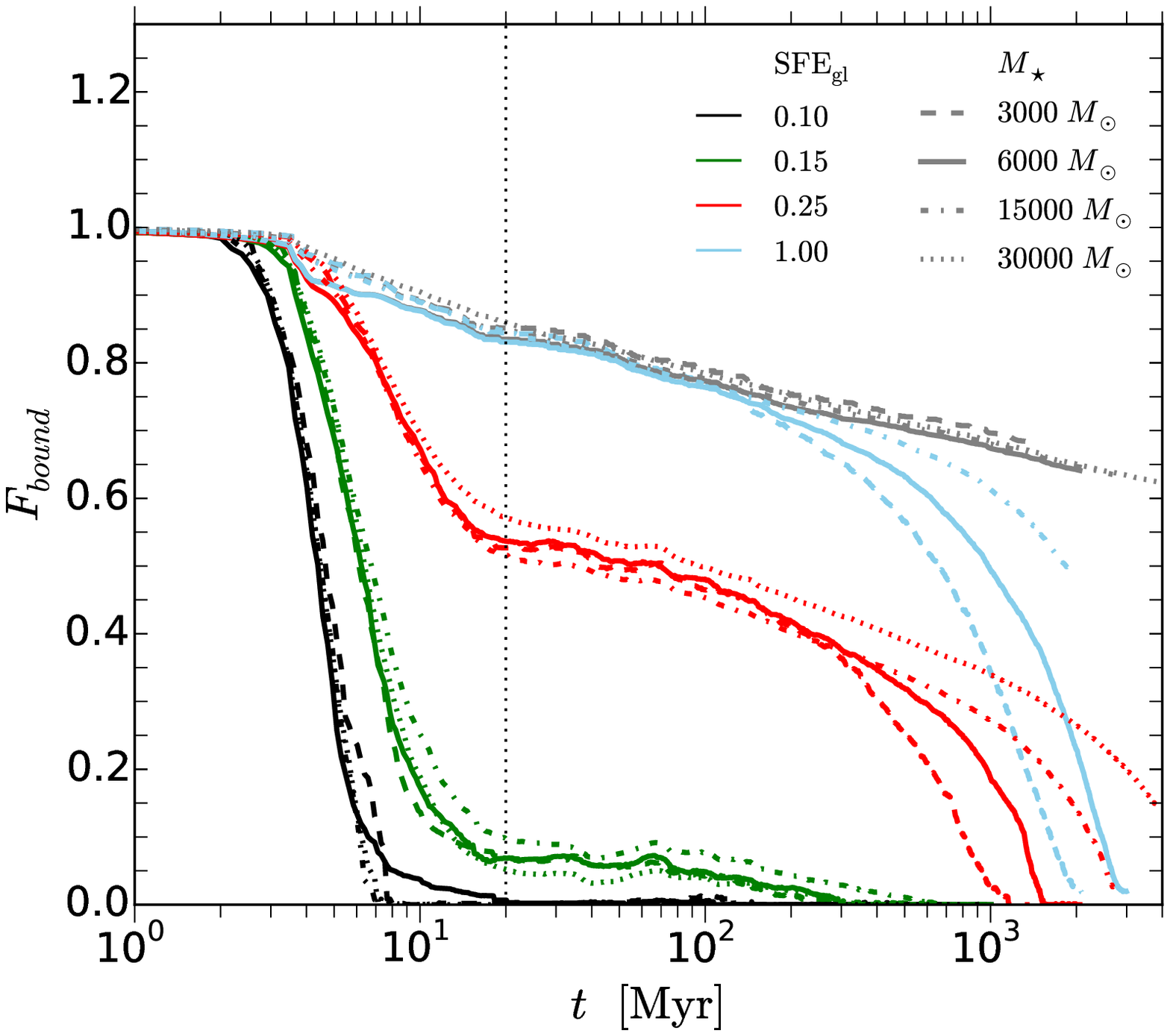}
\caption{Evolution of the bound fraction of clusters obtained in our simulations. Different colors correspond to different global SFEs (see the key). 
The gray lines show the impact of mass loss caused by stellar evolution alone. {The vertical dotted line corresponds to $t = 20$ Myr. }
{Top panel}: comparison of the bound fraction evolution of isolated (dashed lines) and non-isolated (solid lines) models for $M_\star = 6000\ \msun$ with $N\approx10^4$ stars. 
Note that the isolated models are the single-mass models without stellar evolution and scaled to the same physical units as the non-isolated models (i.e., $M_\star = 6000\ \msun$ and $r_h=1.26$~pc).
{Bottom panel}: bound fraction (instantaneous tidal mass as a fraction of initial stellar mass) evolution for non-isolated models with different initial stellar masses (see the key for the line-coding). 
}
\label{fig:tidalmass}
\end{figure}

In the top panel of Fig. \ref{fig:tidalmass} we present the time evolution of the bound fractions of our non-isolated (solid lines) and isolated (dashed lines) models. 
We note that for the isolated clusters we present the final bound fraction defined with the technique described above, and not the fraction of stars with negative total energy. 
This clearly provides a method for determining the bound mass very early 
{(see Appendix \ref{AppenB} ).}
For our non-isolated models, we also show the imprint of stellar-evolution mass loss as the gray lines.
The model clusters with $\mathrm{SFE_{gl}}=1.0$, that is, those that are initially in virial equilibrium without any residual star-forming gas, are depicted as sky-blue lines.
To better visualize both the very fast evolution shortly after gas expulsion and the cluster long-term evolution, the scale of the x-axis (time) is logarithmic.
{The plateau at the very beginning of the bound mass evolution results from our definition of the bound mass fraction, which does not account for the high-velocity unbound stars within the tidal radius.
Because the cluster initial size is smaller than its tidal radius ($r_h/r_t=0.052$), almost all stars reside within the tidal radius during the first few Myr of evolution even when the cluster starts to expand. The bound mass fraction starts to decrease as the escaping stars reach the tidal radius and become unbound by our definition. 
}

{ 
From the bound fraction evolution of our non-isolated model clusters we can identify two regimes of mass loss (the solid lines in the top panel of Fig. \ref{fig:tidalmass}), in addition to the mass loss driven by stellar evolution. 
During the first 20 Myr after gas expulsion, clusters intensively lose their mass (top panel of Fig. \ref{fig:tidalmass}: the solid lines on the left-hand side of the vertical dotted line).
During this time span, the cluster evolution is dominated by the consequences of gas expulsion, and their response is mostly determined by the cluster initial virial ratio.}
{
More or less flat plateaus can be seen in between two identified mass-loss regimes around 20 Myr after gas expulsion.
This means that the surviving part of the cluster is not expanding anymore. }

The bound fraction then decreases more slowly with time. 
It is now mostly affected by stellar evolution and the tidal field of the host galaxy (top panel of Fig. \ref{fig:tidalmass}: the solid lines on the right-hand side of the vertical dotted line).

The bottom panel of Fig. \ref{fig:tidalmass} presents results for the non-isolated models alone for the global SFEs and initial cluster stellar masses quoted in the key. The color-coding is identical to the coding used in the top panel.
We find that the models with identical global SFE show similar evolutionary tracks within the first 20 Myr (i.e., during the violent relaxation) independently of their initial stellar mass (bottom panel of Fig. \ref{fig:tidalmass}). 
In particular, the models with a low global SFE dissolve on similar timescales independently of the initial star cluster mass, $M_\star$ (see the black curves).
Here we recall that the ratio $r_h/r_t$ is kept constant for
now.
The model with a global SFE of 13 percent and $M_\star=6000\ \msun$ does survive as a bound cluster, although with a very small bound fraction, around 2 percent (the blue lines in the
top panel of Fig. \ref{fig:tidalmass}). 
{All other stellar mass models with a global SFE of 13 percent dissolve, however, except for the $M_\star = 30000\ \msun$ model, which barely survives with a 0.17 percent bound fraction, which corresponds to a bound mass of about 52 $\msun$.
Therefore we adopt the models with $\mathrm{SFE_{gl}}$~{=~0.13} as the limit between survival and dissolution following cluster gas expulsion for our adopted tidal field impact $r_h/r_t=0.052$. 
Cluster models with a global SFE of 0.15 and higher can survive instantaneous gas expulsion, as expected from their initial virial ratios (see Table \ref{tab:tsfsfe}).
We note that $\mathrm{SFE_{gl}}$~{=~0.13 is about 2.5} times lower than the SFE threshold for cluster survival when (i) the density profiles of the stars and residual gas have the same shape, (ii) gas expulsion is instantaneous, and (iii) the tidal field impact is negligible. }

\begin{figure}
\centering
\includegraphics[width=\columnwidth]{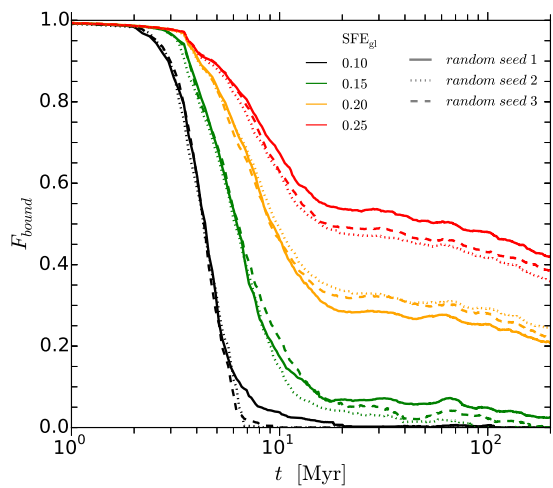}
\caption{ Different random realizations (see line types in the key) of the model clusters with $M_\star=6000~\msun$ ($N=10455$ stars) for $r_h/r_t~=~0.052$.}
\label{fig:rand}
\end{figure}

{We performed a few simulations for a given parameter set but different random seeds to explore the bound fraction variations that are due to random realizations. 
Figure \ref{fig:rand} shows that the bound mass fraction at $t=20$~Myr displays a range of variations of about 10 percent for a cluster with $N=10455$ stars (i.e., $M_\star = 6000\ \msun$ cluster). 
For a cluster with a higher number of stars ($N=26138$, $M_\star = 15000\ \msun$), the range of bound mass fraction variations is about 6 percent. 
The duration of cluster mass-loss in response to gas expulsion remains shorter than 20~Myr for different random realizations.
}

After violent relaxation, the cluster life-expectancy depends on its stellar mass, as expected (red lines in the bottom panel of Fig. \ref{fig:tidalmass}). 
A higher stellar mass implies a higher probability to survive a longer time (but see \citet{Ernst2015}).
The models corresponding to the high global SFEs in this set of simulations show long-term evolutionary patterns similar to the pure Plummer models (i.e., {initially} in virial equilibrium without any external potential).

This study shows us that the mass loss of the cluster in response to gas expulsion is completed within 20 Myr, independently of its initial stellar mass and global SFE, and that its dynamical evolution is mostly affected by the tidal field of the host galaxy thereafter (Fig.~\ref{fig:tidalmass}). 
Since we focus on the cluster bound mass evolution, we consider {the violent relaxation} as the time span when the cluster loses its mass intensively in response to an instantaneous gas expulsion.
We therefore used $t=20$ Myr to measure the final bound fraction of clusters.
We note, however, that the outer shells of surviving clusters need a longer time-span to return to virial equilibrium, as shown by Fig. \ref{fig:lagr-rad} (see below; see also \citet{Brinkmann2017}). 
We note therefore that the violent relaxation duration, as defined here, does not strictly equate with the cluster revirialization time.

An example of the Lagrange radius $R_f$ evolution of a non-isolated cluster with $\mathrm{SFE_{gl}} = 0.20$ and $M_\star=6000 \msun$ is presented in Fig. \ref{fig:lagr-rad}.  
\begin{figure}[!h]
\centering
\includegraphics[width=1.05\columnwidth]{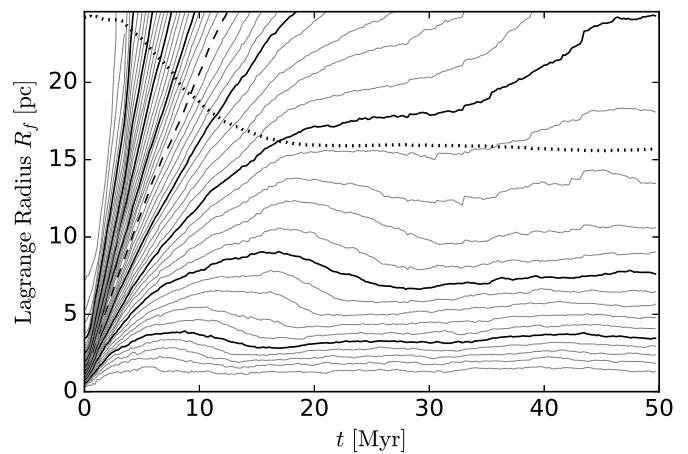}
\caption{ Evolution of a non-isolated cluster of the solar neighborhood  with $\mathrm{SFE_{gl}}=0.20$, $M_\star=6000\ \msun$ and $r_h/r_t=0.052$ over 50 Myr after gas expulsion. We
show 49 Lagrange radii, ranging from 2 percent  to 98 percent  in intervals of 2 percent  (solid and dashed lines). The instantaneous tidal radius $r_t$ of the cluster is shown as the dotted line. Thick solid lines correspond to every 10 percent  of the Lagrange radius, and the dashed line corresponds to 50 percent  of the Lagrange radius.}
\label{fig:lagr-rad}
\end{figure}
The Lagrange radii are defined based on the fraction of initial stellar mass, not on the number fraction of stars. 
The dotted line in this figure represents the instantaneous tidal radius.
This figure shows that the inner parts of a cluster recede, form a bound cluster, and return to virial equilibrium within 20-30 Myr after gas expulsion.
{The inner shells of a cluster revirialize faster than the outer shells, as also found in \citet{Brinkmann2017}, for example.
The cluster tidal radius stays roughly constant after 20 Myr.}

\subsection{Influence of the cluster initial stellar density}

To quantify the effect of the tidal field impact, three additional sets of simulations were performed for cluster masses $M_\star = 6000\ \msun$ and $15000\ \msun$: one with a weaker tidal field impact ($r_h/r_t = $ 0.025), and two with stronger tidal field impact corresponding to $r_h/r_t = $ 0.07 and 0.10. 
That is, we varied the half-mass radius of our model clusters, keeping them in the solar neighborhood (i.e., keeping the same tidal radius for a given stellar mass). 
Because we varied the initial density of the cluster, we varied the normalization factor of time units as shown in Table \ref{tab:tnorm}. 
For denser clusters (smaller $r_h/r_t$), a given physical time-span represents a higher number of N-body time units.

\begin{table}
\centering
\caption{ Scaling factors to the physical time and duration of 20~Myr in N-body units corresponding to different ratios of
the cluster half-mass to tidal radius.}
  \begin{tabular}{ l r r}
  \hline
  \hline
$r_h/r_t$ &
$t_{norm}$  [Myr] &
20 Myr [NB] \\
  \hline

0.025 &
0.0608 &
329 \\
0.052 &
0.1826 &
110 \\
0.07 &
0.2848 &
70 \\
0.10 &
0.4863 &
41 \\
  \end{tabular}
\label{tab:tnorm}
\end{table}

\begin{figure}[!h]
\centering
\includegraphics[width=1.05\columnwidth]{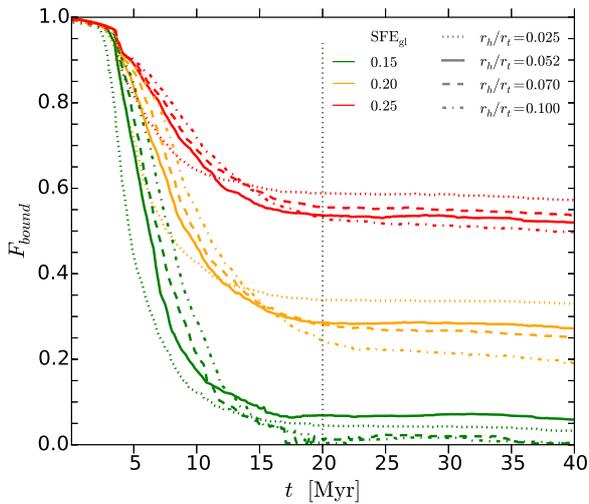}
\caption{  Evolution of the bound fraction of $M_\star = 15000\ \msun$ clusters with different impact of the tidal field (see the key for the line-coding). 
Different colors correspond to different global SFEs (see the key for the color-coding). 
}
\label{fig:Fb_rhrt}
\end{figure}

We present the bound mass fraction evolution of $M_\star$~=~15000~$\msun$ clusters with different initial densities in Fig. \ref{fig:Fb_rhrt}. 
We find that the dense clusters evolve quicker than the less dense clusters within the first 20 Myr after gas expulsion. 
The violent relaxation duration of clusters and their bound mass fraction at $t=20$~Myr depend on the initial cluster mean densities. 
This is consistent with \citet{ParmentierBaumgardt2012} and \citet{Banerjee2013}, who showed that dense clusters have shorter revirialization times than less dense clusters. 
However, we find the cluster violent relaxation to depend fairly weakly on the initial stellar density.
The difference during the first 10 Myr results from our definition of the bound mass fraction.
That  is, the most compact cluster ($r_h/r_t=0.025$) loses mass in response to gas expulsion faster than the most diffuse cluster ($r_h/r_t=0.100$).
Its escaping stars reach the tidal radius twice faster because their velocity dispersions differ by a factor of 2.
Although it is not obvious how to define the duration of violent relaxation accurately, Fig. \ref{fig:Fb_rhrt} shows that it remains shorter than 20-25 Myr regardless of the cluster initial density.
The key to understanding the violent relaxation duration may reside in the mean stellar density within the tidal radius, which
is the same for all considered clusters since they all have the same orbit.

\begin{figure}
\includegraphics[width=\hsize]{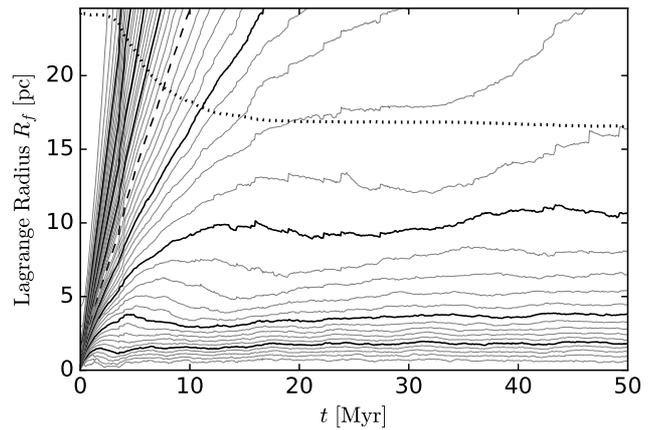}
\caption{ Evolution of the cluster with $M_\star=6000~\msun,~\mathrm{SFE_{gl}}=0.20$ and $r_h/r_t=0.025$ over the first 50 Myr after the instantaneous gas expulsion. }
\label{fig:lagr-rad-compact}
\end{figure}

We note that stars that would be bound to the cluster if the cluster is isolated now become unbound once they cross the tidal radius of the cluster.
Additionally, they are taken away from the cluster by the Galactic tidal field (see the different behaviors of the 30 percent  Lagrange radius in Fig. \ref{fig:lagr-rad} and Fig. \ref{fig:lagr-rad-compact}).

In Fig. \ref{fig:lagr-rad-compact} we present an example of Lagrange radii evolution of a most compact cluster with $M_\star=6000~\msun,~\mathrm{SFE_{gl}}=0.20$. 
We can see that the inner shells of the cluster with $r_h/r_t=0.025$ revirialize faster than those of a more diffuse cluster with $r_h/r_t=0.052$.

Figure~\ref{fig:Fb_rhrt} shows the differences in bound mass fraction at $t=20$~Myr to be around 10 percent between the most compact and the most diffuse clusters. 
More diffuse clusters have a lower bound fraction since their outer shells expand beyond their instantaneous tidal radius.
We have checked that a higher number of particles ($M_\star=15000~\msun$) does not affect our results.

\subsection{Final bound mass fraction in dependence of SFE}
\normalfont

\begin{figure}
\includegraphics[width=\hsize]{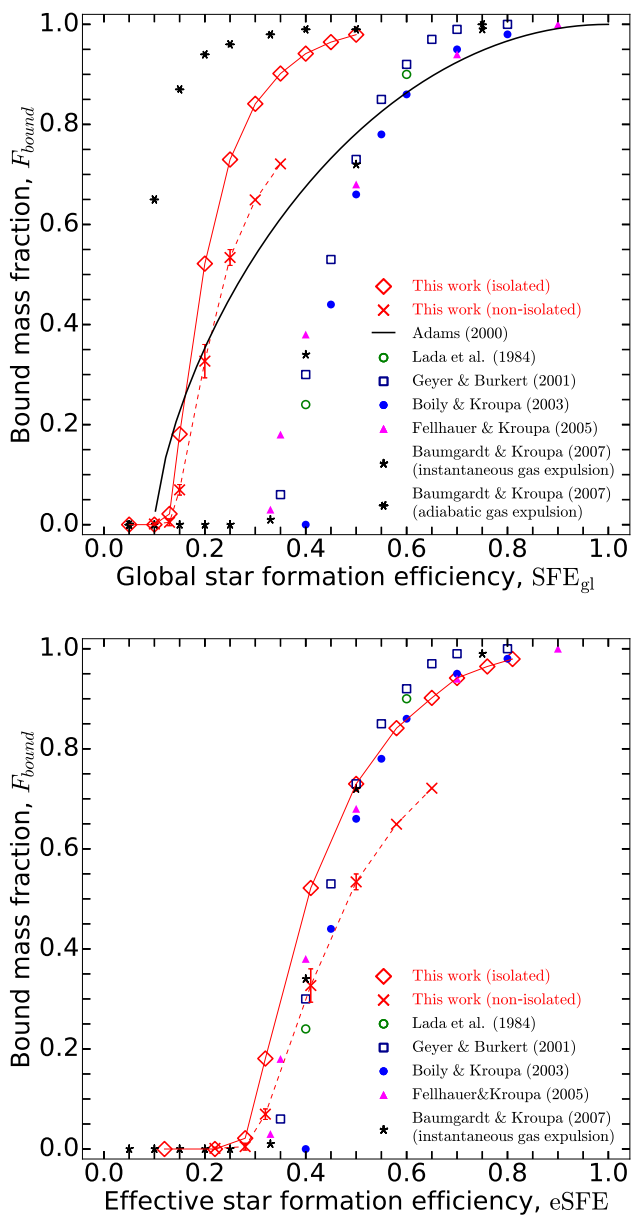}
\caption{Bound fraction as a function of global or effective SFE. We compare our results (red lines) with previous works. The isolated models are depicted by the red diamonds, and non-isolated models by red crosses. In the top panel we use our global SFE ($\mathrm{SFE_{gl}}$) and in the bottom panel, the effective SFE, eSFE$=1/Q_\star$ (which is almost the same as the LSF, the local stellar fraction of our models).}
\label{fig:fb-sfe}
\end{figure}

We compare our results {(represented by the mean values taken from the model clusters with $r_h/r_t = 0.052$ and different stellar masses)} with previous works \citep{Adams2000, Lada1984, Geyer2001, Boily2003b, Fellhauer2005, Baumgardt2007} in the top panel of Fig. \ref{fig:fb-sfe}, which shows the bound fraction as a function of global SFE.
We note here the improved survival likelihood of star clusters after instantaneous gas expulsion, and that although the tidal field is included in our models. 
Our model star clusters whose global SFE is lower than 30 percent, but higher than 15 percent, survive and retain a significant fraction of their stars after violent relaxation.

The global SFEs required by our model clusters to survive gas expulsion provide a good match to the SFEs observed in embedded clusters, which are lower than 30 percent \citep{Higuchi2009,Murray2011,Kainulainen2014}. 
Star clusters are now able to survive instantaneous gas expulsion despite a global SFE as low as 15-20 percent. 
We now have {four} avenues to understand the presence of star clusters with ages older than few Myr in the solar neighborhood despite the low SFE observed: { adiabatic gas expulsion \citep{Lada1984, Geyer2001, Baumgardt2007,Brinkmann2017}, sub-virial clusters \citep{Verschueren1989, Goodwin2009, Farias2015}, hierarchically formed clusters \citep{Smith2011,Lee2016},} and centrally concentrated cluster formation (this contribution).

We argue that the improved survivability of our model clusters is mostly caused by the difference in density profiles between the embedded cluster and its residual gas, as postulated by \citet{PP2013}, namely, the stars have a density profile steeper than that of the residual and initial gas. 
This is the consequence of SF taking place with a constant SFE per free-fall time in a centrally concentrated molecular clump.

We reproduce similar results to previous works when we plot the bound fraction as a function of the effective SFE (eSFE) instead of the global SFE (see the bottom panel of Fig. \ref{fig:fb-sfe}). 
We note here again that in our models the eSFE and the local stellar fraction (LSF) are almost the same.
Our study thus agrees with all present works, including the most recent paper of \citet{Lee2016}, who concluded that the effective SFE is the most important parameter in predicting the cluster survivability.
Our study allows us to compare model SFEs to those achieved in observed forming clusters, however.

\section{Conclusions}\label{sec:conc}

We have performed N-body simulations of violent relaxation and bound cluster formation after instantaneous gas expulsion.
The key point of our study is that we used special initial conditions built on the model of \citet{PP2013}.
This means that the density profile of our model star cluster is steeper than the density profile of the star-forming gas at the time of instantaneous gas expulsion.
If this cluster is in virial equilibrium, including the gravitational potential of residual gas, it should be able to survive gas expulsion despite low SFEs, as shown by \citet{Adams2000} based on a semi-analytical model.

Since our N-body simulations start from the time of instantaneous gas expulsion and do not cover the SF phase, we started with a well-known star cluster model, namely the Plummer model, instead
of starting from the star-free molecular clump.
For this we obtained the dependency of the residual gas density profile on stellar density profile at the time of instantaneous gas expulsion under the assumptions of \citet{PP2013}. In their cluster-formation model, the density profile of a clump (i.e., stars + gas) is constant during SF, and SF takes place with a constant SFE per free-fall time.
Using this, we produced the initial conditions of embedded clusters for our N-body simulations, which depend on the product of two parameters (SFE per free-fall time and SF duration).
With the equations we provide in the Appendix \ref{appenA}, one can use any centrally concentrated spherically symmetric density profile for the embedded cluster and recover the initial and residual gas density profiles.

We adapted the \textsf{falcON} program \textsc{mkhalo} by \citet{McMillan2007} to our problem and have written an additional acceleration plug-in, which represents the gravitational potential of the residual gas in dependence on the SF duration and SFE per free-fall time for a Plummer embedded cluster.
Building on this adapted version of \textsc{mkhalo}, we produced the initial conditions of our simulations, that is, a Plummer star cluster in virial equilibrium with its residual gas with their respective density profiles obeying Eqs. (18) and (19) of \citet{PP2013}.
We related the SF duration to the global SFE to make our set of simulations comparable to earlier works, in which the bound fraction is often presented in dependence on the global SFE.

We performed two types of  cluster simulations, each time covering different global SFEs: 1) isolated single-mass clusters, and 2) non-isolated models, that is, star clusters with stellar evolution and dissolving within a Milky Way-like galaxy. 
We studied the effect of different initial cluster stellar masses as well as of different cluster densities on the evolution of our non-isolated models.
The latter are implemented by varying the cluster half-mass to tidal-radius ratio, $r_h/r_t = 0.025,\ 0.052,\ 0.07,\ \text{and
}0.10$.

Based on the performed simulations, we quantified the bound fraction evolution{ and the violent relaxation duration} of young clusters.
{We defined the violent relaxation duration as the time
span of cluster mass-loss in response to instantaneous gas expulsion.
We note that with our definition, the violent relaxation duration differs from the cluster revirialization time.}
Our models for isolated single-mass clusters allowed us to define an upper limit to the bound fraction as a function of the global SFE.
{For the models considered in our work with their specific parameters -- the stellar density profile (Plummer model), the cluster orbit (with a circular velocity, in the solar neighborhood in the Galactic disk plane), the stellar evolutionary mass-loss from the SSE routine \citep{Hurley2000} and for the cluster mean volume density range ($r_h/r_t=[0.025:0.100] $) -- we conclude that the violent relaxation is not longer than 20 Myr, and its duration depends weakly on the initial stellar density of a cluster.}
We found that the violent relaxation duration of non-isolated model clusters depends neither on the cluster initial stellar mass nor on the global SFE, {keeping the same initial stellar density}.
Varying the tidal field {impact}, that is, varying the cluster size while retaining the cluster mass, does not affect the cluster mass-loss in response to instantaneous gas expulsion much.

We also found that the minimum global SFE necessary to form a bound cluster after instantaneous gas expulsion is $\mathrm{SFE_{gl}}=0.15$ {for a cluster with a circular orbit in the Galactic disk plane at a distance of $R_0 = 8$~kpc from the Galactic center.
If the tidal field is stronger, that is, the cluster is closer to the Galactic center, the minimum $\mathrm{SFE_{gl}}$ needed to survive instantaneous gas expulsion may be different.}
For the given $r_h/r_t$ ratio, the bound fraction of surviving clusters that achieved the same global SFE does not depend on the cluster initial stellar mass.
{The bound mass fraction at the end of violent relaxation for clusters with $r_h/r_t$ of 0.025 and 0.10 differs by only
about 10 percent, with denser clusters retaining a higher bound fraction than more diffuse clusters.}
The evolution of bound clusters after violent relaxation is mostly driven by the tidal field of the host galaxy, and their life expectancy then depends on their stellar mass.

We compared our results with earlier works.
Our final bound fractions are similar to those found in previous works only when the bound fraction is plotted in dependence of the effective SFE.
Thus we agree with \citet{Goodwin2009} that the virial ratio of a cluster at the time of gas expulsion is a key parameter for predicting whether it survives gas expulsion. 
However, when working in terms of the global SFE, that is, the SFE that can be measured by observers as the ratio between the stellar mass and the total (gas+star) mass of a star-forming region, the models proposed in this paper improve the survival likelihood of star clusters after instantaneous gas expulsion.
This is caused by the difference in density profiles between the embedded cluster and its residual gas, namely, the stellar density profile has a steeper slope than that of the residual gas, which is a consequence of SF taking place with a constant star-formation efficiency per free-fall time in a centrally concentrated molecular clump.

\begin{acknowledgements}
{We thank the referee for the helpful comments.}\\
This work was supported by Sonderforschungsbereich SFB 881 "The Milky Way System" (subproject B2) of the German Research Foundation (DFG).\\
The authors gratefully acknowledge Walter Dehnen for his support and discussions in connection with implementing the code \textsc{mkhalo} for our purposes.\\
The authors gratefully acknowledge the computing time granted by the John von Neumann Institute for Computing (NIC) and provided on the supercomputer JURECA at J\"ulich Supercomputing Centre (JSC), under application Nr. 10341 (hhd28).
We also used the smaller GPU cluster KEPLER, funded under the grants I/80 041-043 and
I/81 396 of the Volkswagen Foundation, and the special GPU accelerated supercomputer LAOHU at the Center of Information and Computing at National Astronomical Observatories, Chinese Academy of Sciences, funded by the Ministry of Finance of the People’s Republic of China under the grant ZDYZ2008-2.
\\
BS gratefully acknowledges Rainer Spurzem for providing computing time on the high-performance computing clusters JURECA and LAOHU. 

BS acknowledges the support within program N0073-10/PCF-15-MON by the Ministry of Education and Science of the Republic of Kazakhstan.

PB acknowledges the support by the Chinese Academy of Sciences through the
Silk Road Project at NAOC, through the “Qianren” special foreign experts
program, and the President’s International Fellowship for Visiting
Scientists program of CAS and also the Strategic Priority Research Program
(Pilot B) "Multi-wavelength gravitational wave universe" of the Chinese Academy
of Sciences (No. XDB23040100).

PB acknowledges the support of the Volkswagen Foundation under the Trilateral
Partnerships grant No. 90411 and the special support by the NASU under
the Main Astronomical Observatory GRID/GPU computing cluster project.

\end{acknowledgements}
\normalfont
%
%

   \bibliographystyle{aa} 
   \bibliography{paper1} 

\begin{thebibliography}{}

\bibitem[\protect\citeauthoryear{{Adams}}{{Adams}}{2000}]{Adams2000}
{Adams} F.~C., 2000, \apj, 542, 964

\bibitem[\protect\citeauthoryear{{Banerjee} \& {Kroupa}}{{Banerjee} \&
  {Kroupa}}{2013}]{Banerjee2013}
{Banerjee} S.,  {Kroupa} P., 2013, \apj, 764, 29

\bibitem[\protect\citeauthoryear{{Baumgardt} \& {Kroupa}}{{Baumgardt} \&
  {Kroupa}}{2007}]{Baumgardt2007}
{Baumgardt} H.,  {Kroupa} P., 2007, \mnras, 380, 1589

\bibitem[\protect\citeauthoryear{{Berczik} et~al.}{{Berczik}
  et~al.}{2011}]{Berc2011}
{Berczik} P., {Nitadori} K., {Zhong} S., et~al., 2011, {High performance
  massively parallel direct N-body simulations on large GPU clusters.}, in
  International conference on High Performance Computing, Kyiv, Ukraine,
  October 8-10, 2011., p. 8-18, p.~8

\bibitem[\protect\citeauthoryear{{Boily} \& {Kroupa}}{{Boily} \&
  {Kroupa}}{2003a}]{Boily2003a}
{Boily} C.~M.,  {Kroupa} P., 2003a, \mnras, 338, 665

\bibitem[\protect\citeauthoryear{{Boily} \& {Kroupa}}{{Boily} \&
  {Kroupa}}{2003b}]{Boily2003b}
{Boily} C.~M.,  {Kroupa} P., 2003b, \mnras, 338, 673

\bibitem[\protect\citeauthoryear{{Bonnell} et~al.}{{Bonnell}
  et~al.}{2011}]{Bonnell2011}
{Bonnell} I.~A., {Smith} R.~J., {Clark} P.~C.,  {Bate} M.~R., 2011, \mnras,
  410, 2339

\bibitem[\protect\citeauthoryear{{Brinkmann} et~al.}{{Brinkmann}
  et~al.}{2017}]{Brinkmann2017}
{Brinkmann} N., {Banerjee} S., {Motwani} B.,  {Kroupa} P., 2017, \aap, 600, A49

\bibitem[\protect\citeauthoryear{{Dib} et~al.}{{Dib} et~al.}{2013}]{Dib2013}
{Dib} S., {Gutkin} J., {Brandner} W.,  {Basu} S., 2013, \mnras, 436, 3727

\bibitem[\protect\citeauthoryear{{Duboshin}}{{Duboshin}}{1968}]{Duboshin1968}
{Duboshin} G.~N., 1968, {Nebesnaia mekhanika, Osnovnuye zadachi i metody.}

\bibitem[\protect\citeauthoryear{{Ernst} et~al.}{{Ernst}
  et~al.}{2015}]{Ernst2015}
{Ernst} A., {Berczik} P., {Just} A.,  {Noel} T., 2015, Astronomische
  Nachrichten, 336, 577

\bibitem[\protect\citeauthoryear{{Evans} et~al.}{{Evans}
  et~al.}{2009}]{Evans2009}
{Evans} N.~J., II, {Dunham} M.~M., {J{\o}rgensen} J.~K., et~al., 2009, \apjs,
  181, 321

\bibitem[\protect\citeauthoryear{{Farias} et~al.}{{Farias}
  et~al.}{2015}]{Farias2015}
{Farias} J.~P., {Smith} R., {Fellhauer} M., et~al., 2015, \mnras, 450, 2451

\bibitem[\protect\citeauthoryear{{Fellhauer} \& {Kroupa}}{{Fellhauer} \&
  {Kroupa}}{2005}]{Fellhauer2005}
{Fellhauer} M.,  {Kroupa} P., 2005, \apj, 630, 879

\bibitem[\protect\citeauthoryear{{Fujii} \& {Portegies Zwart}}{{Fujii} \&
  {Portegies Zwart}}{2016}]{Fujii2016}
{Fujii} M.~S.,  {Portegies Zwart} S., 2016, \apj, 817, 4

\bibitem[\protect\citeauthoryear{{Geyer} \& {Burkert}}{{Geyer} \&
  {Burkert}}{2001}]{Geyer2001}
{Geyer} M.~P.,  {Burkert} A., 2001, \mnras, 323, 988

\bibitem[\protect\citeauthoryear{{Girichidis} et~al.}{{Girichidis}
  et~al.}{2012}]{Girichidis2012}
{Girichidis} P., {Federrath} C., {Allison} R., {Banerjee} R.,  {Klessen} R.~S.,
  2012, \mnras, 420, 3264

\bibitem[\protect\citeauthoryear{{Goodwin}}{{Goodwin}}{2009}]{Goodwin2009}
{Goodwin} S.~P., 2009, \apss, 324, 259

\bibitem[\protect\citeauthoryear{{Goodwin} \& {Bastian}}{{Goodwin} \&
  {Bastian}}{2006}]{Goodwin2006}
{Goodwin} S.~P.,  {Bastian} N., 2006, \mnras, 373, 752

\bibitem[\protect\citeauthoryear{{Higuchi} et~al.}{{Higuchi}
  et~al.}{2009}]{Higuchi2009}
{Higuchi} A.~E., {Kurono} Y., {Saito} M.,  {Kawabe} R., 2009, \apj, 705, 468

\bibitem[\protect\citeauthoryear{{Hillenbrand} \& {Carpenter}}{{Hillenbrand} \&
  {Carpenter}}{2000}]{Hillenbrand2000}
{Hillenbrand} L.~A.,  {Carpenter} J.~M., 2000, \apj, 540, 236

\bibitem[\protect\citeauthoryear{{Hills}}{{Hills}}{1980}]{Hills1980}
{Hills} J.~G., 1980, \apj, 235, 986

\bibitem[\protect\citeauthoryear{{Hopkins} et~al.}{{Hopkins}
  et~al.}{2013}]{Hopkins2013}
{Hopkins} P.~F., {Narayanan} D., {Murray} N.,  {Quataert} E., 2013, \mnras,
  433, 69

\bibitem[\protect\citeauthoryear{{Hurley}, {Pols}, \& {Tout}}{{Hurley}
  et~al.}{2000}]{Hurley2000}
{Hurley} J.~R., {Pols} O.~R.,  {Tout} C.~A., 2000, \mnras, 315, 543

\bibitem[\protect\citeauthoryear{{Just} et~al.}{{Just} et~al.}{2009}]{Just2009}
{Just} A., {Berczik} P., {Petrov} M.~I.,  {Ernst} A., 2009, \mnras, 392, 969

\bibitem[\protect\citeauthoryear{{Kainulainen}, {Federrath}, \&
  {Henning}}{{Kainulainen} et~al.}{2014}]{Kainulainen2014}
{Kainulainen} J., {Federrath} C.,  {Henning} T., 2014, Science, 344, 183

\bibitem[\protect\citeauthoryear{{Kroupa}}{{Kroupa}}{2001}]{K2001}
{Kroupa} P., 2001, \mnras, 322, 231

\bibitem[\protect\citeauthoryear{{Krumholz} \& {Matzner}}{{Krumholz} \&
  {Matzner}}{2009}]{Krum2009}
{Krumholz} M.~R.,  {Matzner} C.~D., 2009, \apj, 703, 1352

\bibitem[\protect\citeauthoryear{{Krumholz} \& {Tan}}{{Krumholz} \&
  {Tan}}{2007}]{KrumholzTan2007}
{Krumholz} M.~R.,  {Tan} J.~C., 2007, \apj, 654, 304

\bibitem[\protect\citeauthoryear{{Kudryavtseva} et~al.}{{Kudryavtseva}
  et~al.}{2012}]{Kudryavtseva2012}
{Kudryavtseva} N., {Brandner} W., {Gennaro} M., et~al., 2012, \apjl, 750, L44

\bibitem[\protect\citeauthoryear{{Lada} \& {Lada}}{{Lada} \&
  {Lada}}{2003}]{Lada2003}
{Lada} C.~J.,  {Lada} E.~A., 2003, \araa, 41, 57

\bibitem[\protect\citeauthoryear{{Lada}, {Lombardi}, \& {Alves}}{{Lada}
  et~al.}{2010}]{Lada2010}
{Lada} C.~J., {Lombardi} M.,  {Alves} J.~F., 2010, \apj, 724, 687

\bibitem[\protect\citeauthoryear{{Lada}, {Margulis}, \& {Dearborn}}{{Lada}
  et~al.}{1984}]{Lada1984}
{Lada} C.~J., {Margulis} M.,  {Dearborn} D., 1984, \apj, 285, 141

\bibitem[\protect\citeauthoryear{{Lada} et~al.}{{Lada} et~al.}{1991}]{Lada1991}
{Lada} E.~A., {Depoy} D.~L., {Evans} N.~J., II,  {Gatley} I., 1991, \apj, 371,
  171

\bibitem[\protect\citeauthoryear{{Lee} \& {Goodwin}}{{Lee} \&
  {Goodwin}}{2016}]{Lee2016}
{Lee} P.~L.,  {Goodwin} S.~P., 2016, \mnras, 460, 2997

\bibitem[\protect\citeauthoryear{{Leisawitz}, {Bash}, \&
  {Thaddeus}}{{Leisawitz} et~al.}{1989}]{Leisawitz1989}
{Leisawitz} D., {Bash} F.~N.,  {Thaddeus} P., 1989, \apjs, 70, 731

\bibitem[\protect\citeauthoryear{{McMillan} \& {Dehnen}}{{McMillan} \&
  {Dehnen}}{2007}]{McMillan2007}
{McMillan} P.~J.,  {Dehnen} W., 2007, \mnras, 378, 541

\bibitem[\protect\citeauthoryear{{Miyamoto} \& {Nagai}}{{Miyamoto} \&
  {Nagai}}{1975}]{Miyamoto1975}
{Miyamoto} M.,  {Nagai} R., 1975, \pasj, 27, 533

\bibitem[\protect\citeauthoryear{{Moeckel} et~al.}{{Moeckel}
  et~al.}{2012}]{Moeckel2012}
{Moeckel} N., {Holland} C., {Clarke} C.~J.,  {Bonnell} I.~A., 2012, \mnras,
  425, 450

\bibitem[\protect\citeauthoryear{{Murray}}{{Murray}}{2011}]{Murray2011}
{Murray} N., 2011, \apj, 729, 133

\bibitem[\protect\citeauthoryear{{Murray}, {Quataert}, \& {Thompson}}{{Murray}
  et~al.}{2010}]{Murray2010}
{Murray} N., {Quataert} E.,  {Thompson} T.~A., 2010, \apj, 709, 191

\bibitem[\protect\citeauthoryear{{Parmentier} \& {Baumgardt}}{{Parmentier} \&
  {Baumgardt}}{2012}]{ParmentierBaumgardt2012}
{Parmentier} G.,  {Baumgardt} H., 2012, \mnras, 427, 1940

\bibitem[\protect\citeauthoryear{{Parmentier} \& {Gilmore}}{{Parmentier} \&
  {Gilmore}}{2007}]{Parmentier2007}
{Parmentier} G.,  {Gilmore} G., 2007, \mnras, 377, 352

\bibitem[\protect\citeauthoryear{{Parmentier} \& {Pfalzner}}{{Parmentier} \&
  {Pfalzner}}{2013}]{PP2013}
{Parmentier} G.,  {Pfalzner} S., 2013, \aap, 549, A132

\bibitem[\protect\citeauthoryear{{Pfalzner} et~al.}{{Pfalzner}
  et~al.}{2014}]{PP2014}
{Pfalzner} S., {Parmentier} G., {Steinhausen} M., {Vincke} K.,  {Menten} K.,
  2014, \apj, 794, 147

\bibitem[\protect\citeauthoryear{{Plummer}}{{Plummer}}{1911}]{Plummer1911}
{Plummer} H.~C., 1911, \mnras, 71, 460

\bibitem[\protect\citeauthoryear{{Proszkow} \& {Adams}}{{Proszkow} \&
  {Adams}}{2009}]{Proszkow2009}
{Proszkow} E.-M.,  {Adams} F.~C., 2009, \apjs, 185, 486

\bibitem[\protect\citeauthoryear{{Reggiani} et~al.}{{Reggiani}
  et~al.}{2011}]{Reggiani2011}
{Reggiani} M., {Robberto} M., {Da Rio} N., et~al., 2011, \aap, 534, A83

\bibitem[\protect\citeauthoryear{{Smith} et~al.}{{Smith}
  et~al.}{2011}]{Smith2011}
{Smith} R., {Fellhauer} M., {Goodwin} S.,  {Assmann} P., 2011, \mnras, 414,
  3036

\bibitem[\protect\citeauthoryear{{Smith} et~al.}{{Smith}
  et~al.}{2013}]{Smith2013}
{Smith} R., {Goodwin} S., {Fellhauer} M.,  {Assmann} P., 2013, \mnras, 428,
  1303

\bibitem[\protect\citeauthoryear{{Tutukov}}{{Tutukov}}{1978}]{Tutukov1978}
{Tutukov} A.~V., 1978, \aap, 70, 57

\bibitem[\protect\citeauthoryear{{Verschueren} \& {David}}{{Verschueren} \&
  {David}}{1989}]{Verschueren1989}
{Verschueren} W.,  {David} M., 1989, \aap, 219, 105

\end{thebibliography}
\begin{appendix} 
\section{Expression of the density profile of unprocessed gas}\label{appenA}
\normalfont
Equation (\ref{polynom}) can be easily solved using software
such as \textsc{mathematica}. Since the roots of this equation obtained with \textsc{mathematica} are very long, we made them more compact by introducing the following intermediate terms:
\begin{equation}
\alpha = k^{4}\rho_\star^2\ ;
\end{equation}
\begin{equation}
K_0=\sqrt[3]{\alpha^3+36 \alpha^2+216 \alpha +24 \alpha \sqrt{3 \left(\alpha+27\right)}}\ ;\quad \\
\end{equation}
\begin{equation}
K_1=\sqrt{\frac{\alpha^2+\alpha(K_0+24)+K_0 (K_0+12)}{12 k^4 K_0}};\quad \\
\end{equation}
\begin{equation}
K_2=\frac{\left(\alpha-K_0+24\right) \left(K_0-\alpha\right)}{3 k^4 K_0}.\quad
\end{equation}
Then we can write the four roots of Eq. (\ref{polynom}) as
\begin{equation}
\label{roots}
{\rho_\mathrm{gas(1,2,3,4)}}=\frac{1}{k^2}-\frac{\rho_\star}{2} \pm \frac{1}{2} \sqrt{K_2 + \frac{8}{k^6 (\mp K_1)}} + (\mp K_1).\\
\end{equation}
The following relation is true for all real values of $k$ and $\rho_\star$
\begin{equation}
\label{sroot}
K_2 < \frac{8}{k^6 K_1}
,\end{equation}
which gives complex numbers for two of the roots in case of $(-K_1)$.
The other two roots are real, with one decreasing and the other increasing with $\rho_\star$. 

In our case, we expect the residual gas density profile to be a decreasing function of the radius, as the stellar density profile. 
Thus we choose the root that increases together with stellar density toward the clump center with the following expression:
\begin{equation}
\label{rootsf}
{\rho_\mathrm{gas}}=\frac{1}{k^2}-\frac{\rho_\star}{2} - \frac{1}{2} \sqrt{K_2 + \frac{8}{k^6 K_1}} + K_1.\\
\end{equation}

\section{Bound fraction of isolated models}\label{AppenB}

\begin{figure}[!h]
\centering
\includegraphics[width=1.05\columnwidth]{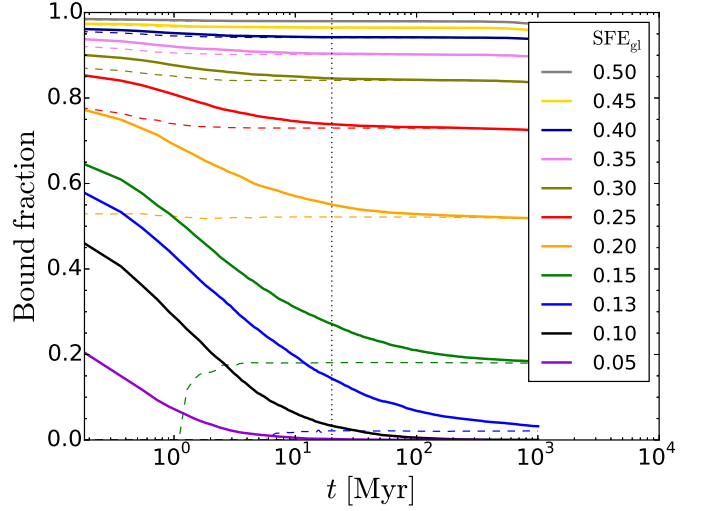}
\caption{
{Time evolution of the bound fraction $F_{b}$ of isolated models ($N=10^4$) as defined by two methods: defined by the fraction of stars with a negative total energy (solid lines), and defined by recalculating the total energy of stars in an iterative process (see text for details; dashed lines). The vertical dotted line corresponds to $t = 20$~Myr when we scale the isolated models with the same scale factor as for a non-isolated model with $M_\star=6000\ \msun$, which also has $N\approx10^4$.}}
\label{fig:Fbnd}
\end{figure}

For isolated models we first defined the bound fraction based on the total (i.e., kinetic + potential) energy of stars, that
is, as the fraction of stars with a negative total energy (solid lines in Fig. \ref{fig:Fbnd}).
Since our model clusters become super-virial after instantaneous gas expulsion, the unbound stars can be located anywhere inside the cluster and make a significant contribution to its gravitational potential depending on global SFE. 
Thus the bound fraction remains overestimated until bound and unbound stars are clearly spatially separated from each other.
For instance, model clusters with global SFEs of 0.05 and 0.10 do not survive the instantaneous gas expulsion. In Fig. \ref{fig:Fbnd}, however, they retain a significant bound fraction for several Myr.
For clusters with global SFEs of 0.13 and 0.15, we need to wait for a long time to reach the final bound fraction, that is, for the unbound stars to have evacuated the cluster region and to no longer contribute to the cluster gravitational potential.

These reasons motivated us to develop another technique to define the final bound fraction early on in the evolution of clusters.
To do so, we recalculated the total energy of each star after removing the unbound stars. 
Some of these recalculated energies are now positive (i.e., some previously bound stars are now unbound) since the cluster gravitational potential is now shallower. 
We removed this new sample of unbound stars, and continued to iterate until only bound stars were left within our selection.
The fraction of stars left in this selection determines the final bound fraction.

\begin{figure}
\includegraphics[width=\hsize]{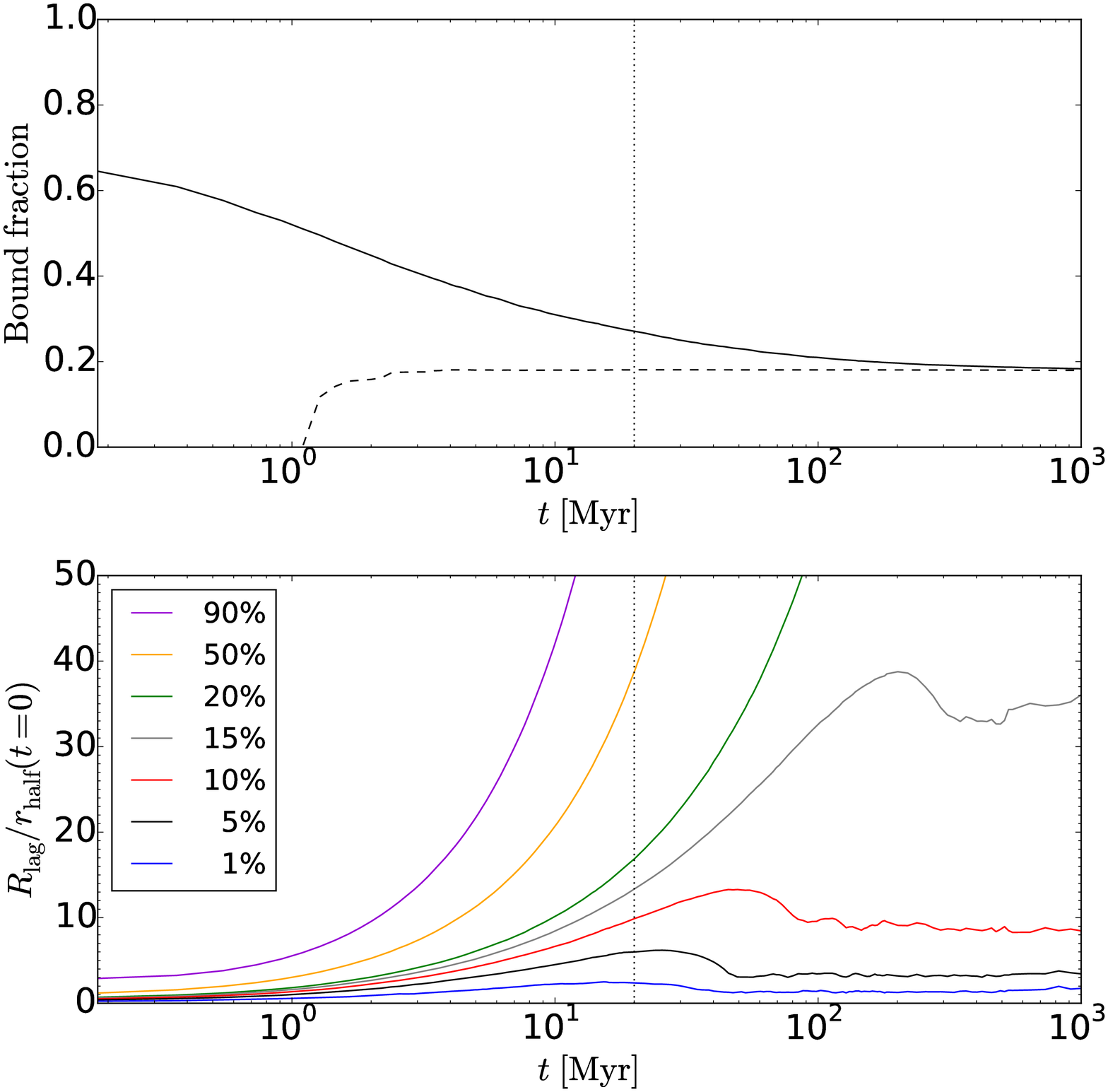}
\caption{Top panel: bound and final bound fraction evolution of an isolated cluster with $\mathrm{SFE_{gl}}=0.15$ (solid and dashed lines, respectively).
Bottom panel: Lagrange radius evolution of the same model. Lagrange radii are given in units of the initial stellar half-mass radius.
}
\label{fig:lagrb}
\end{figure}

Figure \ref{fig:lagrb} presents an example of the evolution of the bound fraction and Lagrange radii of an isolated cluster with a global SFE of 15 percent.
As we see when comparing the top and bottom panels, the bound fraction decreases during the cluster expansion and reaches its final value when its bound stars have collapsed back and form a bound cluster. 
With the new technique, however, we can predict the final bound fraction already after 2 Myr, when the cluster is still in the
expansion phase.
  Figure \ref{fig:Fbnd} and the top panel of Fig. \ref{fig:lagrb} show that the instantaneous bound fraction converges toward the final bound fraction determined with our technique by the end of the simulations. 
This shows that with our calculation method we can estimate the final bound fraction even before the inner part of the cluster starts to collapse back and return to virial equilibrium. 
We caution, however, that with this method we underestimate the final bound fraction of a cluster with a low global SFE during their early evolution after instantaneous gas expulsion. 
This is caused by removing all unbound stars, including the centrally
concentrated ones, which contribute the most to the gravitational field of the cluster (see Fig. \ref{fig:lagrb}). 
This is the reason for the unusual behavior of the final bound fractions of isolated clusters with a global SFE of 0.13 and 0.15, which is 0 at $t \lapprox 1$~Myr, and why  they rise at an early time in the evolution instead of decreasing.

\end{appendix}
\end{document}